# PROTECTED GROUNDS AND THE SYSTEM OF NON-DISCRIMINATION LAW IN THE CONTEXT OF ALGORITHMIC DECISION-MAKING AND ARTIFICIAL INTELLIGENCE

PROF. DR. JANNEKE GERARDS & PROF. DR. FREDERIK ZUIDERVEEN BORGESIUS†

*Algorithmic decision-making and similar types of artificial intelligence (AI) may lead to improvements in all sectors of society but can also have discriminatory effects. While current non-discrimination law offers people some protection, algorithmic decision-making presents the lawmakers and law enforcement with several challenges. For instance, algorithms can generate new categories of people based on seemingly innocuous characteristics, such as web browser preference or apartment number, or more complicated categories combining many data points. Such new types of differentiation could evade non-discrimination law, as browser type and house number are not protected characteristics, but such differentiation could still be unfair, for instance if it reinforces social inequality.*

*This paper attempts to determine which system of non-discrimination law would best be applied to algorithmic decision-making, considering that algorithms can differentiate on the basis of characteristics that do not correlate with protected grounds of discrimination such as ethnicity or gender. The paper analyses the current loopholes in the protection offered by non-discrimination law and explores the best way for lawmakers to approach the issue of algorithmic differentiation. While we focus on Europe, this paper's concentration on concept and theory rather than specific application, should prove useful for scholars and policymakers from*

† Prof. Dr. Janneke Gerards is full professor of fundamental rights law at Utrecht University, the Netherlands. Prof. Dr. Frederik Zuiderveen Borgesius is full professor ICT and law at the Interdisciplinary research hub on Security, Privacy, and Data Governance (iHub) and the Institute for Computing and Information Sciences (iCIS), Radboud University Nijmegen, The Netherlands. We would like to thank the participants to our session at the Privacy Law Scholars Conference (PLSC), June 4-5, 2020, Washington, United States, in particular Sam Wrigley and Deven Desai. We would also like to thank Marvin van Bekkum and Florianne Peters van Neijenhof for excellent research assistance, and Laurens Naudts for his useful suggestions.

*other regions as they encounter similar problems with algorithmic decision-making.*

INTRODUCTION

Algorithmic decision-making and artificial intelligence ("AI") may lead to improvements in all sectors of society but can also have discriminatory effects. While current non-discrimination law offers people some protection, AI decision-making presents the law with several challenges.

For instance, algorithms can generate new categories of people based on seemingly innocuous characteristics, such as web browser preference or apartment number, or more complicated categories combining many data points. To illustrate: Dutch insurance companies charged extra for car insurance if customers lived in apartments with certain types of numbers (a letter included in the number, such as 4A or 20C).[1] As another example: an online store may find that most consumers using a certain web browser pay less attention to prices; the store can charge those consumers extra.

Such new types of differentiation could evade non-discrimination law, as browser type and house number are not protected characteristics, but such differentiation may still be unfair, for example because it could reinforce social inequality. In this light, this paper examines the following issue: What system of non-discrimination law can best be applied to algorithmic decision-making, knowing that algorithms can differentiate on the basis of characteristics that do not correlate with protected grounds such as ethnicity or gender?

The paper explores the best way for lawmakers to approach algorithmic differentiation based on non-protected grounds. The paper also discusses what system of non-discrimination law is the best instrument to mitigate risks of algorithmic differentiation. While we focus on Europe, the conceptual and theoretical focus of the paper can make it useful for scholars and policymakers from other regions, as they encounter similar problems with algorithmic decision-making.

---

    1. Gerald Koon, *Premie autoverzekering op basis regio en postcode [Premiums for Car Insurance Based on Region and Postal Code]*, CONSUMENTENBOND [DUTCH CONSUMER ORGANISATION] (Nov. 29, 2017), https://www.consumentenbond.nl/autoverzekering/je-postcode-en-de-premie.

The paper contributes to the debate on algorithmic differentiation and non-discrimination law, both in Europe and beyond. The paper differs from existing scholarship in several ways. First, the paper combines two fields of law (non-discrimination law and information and communications technology law) and was written by a scholar in each of those respective fields. Second, the paper applies theories from non-discrimination scholarship, for instance on open and closed lists of protected grounds, to problems posed by algorithmic differentiation. This is among the first papers to analyze, in-depth, the issue of protected grounds in European non-discrimination law in the context of algorithmic differentiation.[2]

In this paper, we usually speak of discrimination when discrimination harms people with protected characteristics (such as ethnicity), constitutes unjustified unequal treatment, or is generally considered objectionable. We speak of differentiation, making distinctions/classifications, or unequal treatment when we refer to discrimination in a neutral sense. When discussing algorithmic decision-making, we refer to decisions made on the basis of the output of a computer algorithm. (See Section I for more details.)

To keep the discussion manageable, we focus on examples taken from the private sector (e.g. employment, provision of goods and services). This paper focuses on non-discrimination law, and does not discuss, for instance, data protection law[3] or consumer protection law as such.[4] The paper further focuses on one problem with

---

2. Alessandro Mantelero, *From Group Privacy to Collective Privacy: Towards a New Dimension of Privacy and Data Protection*, GRP. PRIV.: NEW CHALLENGES OF DATA TECH. 173 (Linnet Taylor et al. eds., 2017); Brent Mittelstadt, *From Individual to Group Privacy in Big Data Analytics*, 30 PHIL. & TECH. 475 (2017); Sandra Wachter, *Affinity Profiling and Discrimination by Association in Online Behavioural Advertising*, 35 BERKELEY TECH. L.J. 367 (2020); Laurens Naudts, *Criminal Profiling and Non-Discrimination: On Firm Grounds for the Digital Era?*, *in* SEC. & L. LEGAL & ETHICAL ASPECTS OF PUB. SEC., CYBER SEC. & CRITICAL INFASTRUCTURE SEC. (Anton Vedder et al. eds., 2019); Tal Z. Zarsky, *An Analytic Challenge: Discrimination Theory in the Age of Predictive Analytics*, 14 I/S J. L. & POL'Y FOR INFO. SOC'Y 11 (2017).

3. *See* Philipp Hacker, *Teaching Fairness to Artificial Intelligence: Existing and Novel Strategies Against Algorithmic Discrimination Under EU Law*, 55 COMMON MKT. L. REV. 1143 (2018); Sandra Wachter & Brent Mittelstadt, *A Right to Reasonable Inferences: Re-Thinking Data Protection Law in the Age of Big Data and AI*, 2019 COLUM. BUS. L. REV. 494 (2019).

4. *See* Agnieszka Jabłonowska et al., *Consumer Law and Artificial Intelligence. Challenges to the EU Consumer Law and Policy Stemming from the Business' Use of Artificial Intelligence. Final Report of the ARTSY Project* (Eur. Univ. Inst. L. Working Paper, Nov. 11, 2018), https://cadmus.eui.eu/handle/1814/57484; Frederik J. Zuiderveen Borgesius, *Strengthening Legal Protection Against Discrimination by* non-discrimination law: the concept of protected grounds. Issues such as enforcement of non-discrimination law and discrimination-aware AI are outside the scope of this paper.[5]

The paper concentrates on non-discrimination law derived from the Council of Europe (with an emphasis on the European Convention on Human Rights) and the EU (the Charter of Fundamental Rights of the European Union and secondary EU non-discrimination legislation). National law is outside the scope of this paper. The EU's General Data Protection Regulation is also outside the scope of this paper (except for some brief remarks in Section II).

The paper is structured as follows. Section I introduces algorithmic decision-making and some related concepts. We distinguish two categories of risks related to algorithmic decision-making: discrimination and unfair differentiation. Section II introduces non-discrimination law. Section IV analyzes the difference between closed and open lists of protected grounds in non-discrimination law. Section V analyzes the characteristics, strengths and weaknesses of open and closed lists in non-discrimination law. Section VI examines what non-discrimination system would be best suited to dealing with the challenges of algorithmic discrimination.

I. ALGORITHMIC DISCRIMINATION AND DIFFERENTIATION

*A. Algorithms, Artificial Intelligence, and Machine Learning*

This Section introduces algorithmic decision-making. An algorithm is "an abstract, formalized description of a computational procedure."[6] In this paper, we call the outcome of such a procedure a decision. We focus on decisions that affect people. Loosely speaking, one can think of an algorithm as a computer program.

A concept closely related to algorithmic decision-making is artificial intelligence. AI can be described as "[t]he theory and

---

*Algorithms and Artificial Intelligence*, 24 INT'L J. HUM. RTS. 1572 (Mar. 25, 2020), https://doi.org/10.1080/13642987.2020.1743976.

5. *See* Robin Allen & Dee Masters, *Regulating for an Equal AI: A New Role for Equality Bodies Meeting the New Challenges to Equality and Non-Discrimination from Increased Digitisation and the use of Artificial Intelligence*, EQUINET 2020, https://equineteurope.org/wp-content/uploads/2020/06/ai_report_digital.pdf [https://perma.cc/3B3M-F5U3]; *See AI: ACM Conference on Fairness, Accountability, and Transparencey (ACM FACCT)*, ASSOCIATION FOR COMPUTING MACHINERY, https://facctconference.org [https://perma.cc/2BHR-M49T] (last visited Sept. 21, 2021) [hereinafter ACM CONFERENCE].

6. Paul Dourish, *Algorithms and Their Others: Algorithmic Culture in Context*, BIG DATA & SOC'Y, July–Dec. 2016, 1, 3; *see also* Deven R. Desai & Joshua A. Kroll, *Trust but Verify: A Guide to Algorithms and the Law*, 31 HARV. J.L. & TECH. 1, 1 (2017) [hereinafter Desai & Kroll].

development of computer systems able to perform tasks normally requiring human intelligence."[7] AI can be divided into two broad categories: (i) rule-based AI, and (ii) machine learning.[8] For rule-based, or knowledge-based, AI, usually "explicit facts and rules about some activity are explicitly programmed into software, harnessing the knowledge of domain experts about how some system or activity operates."[9]

Machine learning can be described as "an automated process of discovering correlations (sometimes alternatively referred to as relationships or patterns) between variables in a dataset, often to make predictions or estimates of some outcome."[10] One could also say that machine learning is "statistics on steroids".[11] Machine learning has become the most successful type of AI. In fact, machine learning is so successful that AI is often used as a synonym for machine learning now. For ease of reading, we will mostly speak of AI in this paper.

Algorithms are good at finding correlations in data sets. Suppose that you want to develop a spam filter. You start with a set of training data. The set consists of a million email messages, of which 60% has been labelled as "spam" by people in the past. The other 40% has not been labelled spam and can be assumed to be non-spammy emails. The algorithmic system can find correlations between the contents of messages and whether they were labeled "spam". The system might learn, for instance, to recognize spam messages on the basis of characteristics such as their length; the words included in the message; spelling errors; the sender, etc. Based on such characteristics, the algorithm can differentiate between spam and not-spam when the system encounters new messages (that were not part of the training data).

Machine learning often involves constructing a predictive model. In a simplified example, such a model could look like this: *"If an email message has characteristics A, B, C, and D, then there is a 99% chance that a human would label it as spam."*

Algorithms can also be used to make decisions about people. For instance, a bank could use an algorithm to predict whether

---

people will pay back loans. The AI system can be trained with data about repaying loans from the past. A predictive model could say: "If an individual has characteristics E. F, G, and H, then there is a 5% chance that the person will not pay back a loan." The bank could use the algorithm to assess the chance that a new customer (who provides certain data about him or herself) will repay a loan. The bank could even let the algorithm decide, automatically, whether people are granted a loan. In the remainder of this paper, we sacrifice some precision for readability when mentioning technology. We speak of algorithmic decision-making; that phrase may refer both to different types of AI and simpler automated systems.

## B.  *Algorithmic Decision-making and Harm to Protected Groups*

Algorithmic decision-making could give an impression of rationality, efficiency, and infallibility; numbers and mathematics do not lie. Some might say that the algorithms in themselves are neutral.[12] Unfortunately, however, as we will see in this paper, algorithmic decision-making can also lead to discrimination on the basis of, for instance, ethnicity or gender, or could lead to other forms of unfair differentiation.

Algorithmic differentiation presents non-discrimination law with several challenges. We focus on two categories of problems: (i) differentiation that harms people with protected characteristics, such as ethnicity or gender, and (ii) differentiation that does not harm people with protected characteristics, but that is still unfair.[13]

We start with problem (i), where people with protected characteristics are harmed. Such unintentional algorithmic discrimination can happen, for instance, if an algorithmic application is trained on a data set that incorporated discriminatory or biased human decisions.[14] A new subfield in computer science—fairness and transparency in machine learning—focuses on uncovering and mitigating unfairness and discrimination in algorithms.[15] Computer scientists, other academics, and journalists have uncovered many examples of algorithmic decision-making with discriminatory effects.[16] Below, we give a few examples of how such decision-making could harm protected groups.

---

12. *Cf* Gillian Tett, *Mapping Crime – or Stirring Hate? When Chicago Police Ran Their Predictive Statistics, There Was a Strong Racial Imbalance*, FIN. TIMES (Aug. 22, 2014), https://www.ft.com/content/200bebee-28b9-11e4-8bda-00144feabdc0.

13. *See* Tal Zarsky, *The Trouble with Algorithmic Decisions: An Analytic Road Map to Examine Efficiency and Fairness in Automated and Opaque Decision Making*, 41 SCI. TECH. & HUM. VALUES 118 (2016) [hereinafter Zarsky].

14. Discriminatory training data are merely one cause of discriminatory AI. For other causes of discriminatory AI, *see* Solon Barocas & Andrew D. Selbst, *Big Data's Disparate Impact*, 104 CALIF. L. REV. 671 (2016).

15. *Fairness, Accountability, And Transparency In Machine Learning*, https://www.fatml.org (last visited Sept. 21, 2021); *see also* ACM CONFERENCE, *supra* note 5.

A real-life example from the UK concerns admissions to a medical university in the 1980s.[17] To help the school deal with the many applications it received each year, an employee developed a computer system to select the best students. The employee made the computer system reproduce the decisions by human selectors from earlier years. As the Commission for Racial Equality observed,

> His aim was to remove any inconsistencies in the selection. In devising his program he took account, not only of academic performance, but of all criteria considered relevant in the selection of students. He did this by observing the decisions of the selectors over a number of years and adjusting the program accordingly. In short, the program aimed to *mimic* the judgements of the selectors.[18]

To make the system's decisions resemble the human decisions from earlier years, the employee included negative weightings for both female and non-Caucasian in the computer program. The system rejected a disproportionate number of female and non-white applicants.

The Commission for Racial Equality noted that the "computer program (…) replicated the discrimination that was already being practised by the selectors."[19] The data in previous years reflected (conscious or subconscious) discriminatory decisions by humans. The Commission concluded that the university illegally discriminated with the computer system.[20]

In the previous example, the computer programmer consciously added gender and ethnicity to a program to mimic human decisions. But nowadays, algorithmic systems often find correlations in training data in an automated way. If an algorithmic system is trained on data that reflects discriminatory decisions from the past, the system risks reproducing that bias.[21]

---

16. *See, e.g.*, Desai & Kroll, *supra* note 6, at 1; Frederik J. Zuiderveen Borgesius, *Strengthening Legal Protection Against Discrimination by Algorithms and Artificial Intelligence*, 24 INTL J. HUM. RTS. 1572 (Mar. 25, 2020), https://www.tandfonline.com/doi/pdf/10.1080/13642987.2020.1743976?needAccess=true [hereinafter *Zuiderveen Borgesius*]; CATHY O'NEIL, WEAPONS OF MATH DESTRUCTION: HOW BIG DATA INCREASES INEQUALITY AND THREATENS DEMOCRACY (2016).

17. COMM'N FOR RACIAL EQUAL., MEDICAL SCHOOL ADMISSIONS: REPORT OF A FORMAL INVESTIGATION INTO ST. GEORGE'S HOSPITAL MEDICAL SCHOOL (1988) [hereinafter COMM'N FOR RACIAL EQUALITY]; *see also* Stella Lowry & Gordon Macpherson, *A Blot on the Profession*, 296 BRIT. MED. J. (CLINICAL RSCH. ED.) 657 (1988); Solon Barocas & Andrew D. Selbst, *Big Data's Disparate Impact*, 104 CALIF. L. REV. 671, 682 (2016) [hereinafter Barocas & Selbst].

18. COMM'N FOR RACIAL EQUALITY, *supra* note 17, at 8.

19. *Id.* at 15.

20. *Id.* at 11.

To illustrate, in 2013, it was shown that Google's system for targeting advertising to search engine users was influenced by biased human behavior.[22] When people searched for African-American-sounding names in the Google search engine, Google showed advertisements that suggested that somebody had an arrest record.[23] When people searched for white-sounding names, Google displayed fewer ads suggestive of arrest records.[24] Presumably, Google's algorithm analyzed which ads were clicked on the most, and thus inherited a racial bias.[25] Willful algorithmic discrimination is also possible. A company using algorithms as a basis (or bases) for their decisions could discriminate on purpose and hide that discrimination by discriminating by proxy ("masking"). To give a simplified example, the company could find a correlation between living in a certain postal code and ethnicity. If a company wanted to discriminate against job applicants with a certain ethnicity, the company could try to hide that discrimination by discriminating on a proxy: postal code.[26]

Using algorithms, the company could find more complicated proxies. For instance, a company might discover that a certain ethnicity correlates with a combination of data. Say a correlation is as follows: 90% of the following group has a dark skin color: people who drive a car of the model X of the brand Y, live in an area with postal code ABCD, and are born between 1981 and 1987. We will call this: "Proxy Group 1"

Hence, a company that wants to discriminate against dark-skinned people could discriminate against that group, by discriminating against Proxy Group 1. As Custers et al. note, "novel predictive models can prove to be no more than sophisticated tools to mask the 'classic' forms of discrimination of the past, by hiding it behind scientific findings and faceless processes."[27] In sum, a company could discriminate, on purpose, against people with a certain ethnicity, but could hide that discrimination behind proxies.

---

21. *See* Barocas & Selbst, *supra* note 17, 681–85.
22. Latanya Sweeney, *Discrimination in Online Ad Delivery*, 56 COMMC'NS. ACM 44, 47 (May 2013).
23. *Id*.
24. *Id*.
25. *Id*. at 48.
26. *See* Barocas & Selbst, *supra* note 17, at 712.
27. Bart Custers et al., *The Way Forward*, *in* 3 STUDIES IN APPLIED PHIL., EPISTEMOLOGY AND RATIONAL ETHICS. DISCRIMINATION & PRIV. IN THE INFO. SOC'Y 341, 352 (Bart Custers et al. eds., Springer 2013) [hereinafter Custers].

As noted, accidental discrimination by proxy is also possible, and probably happens more often than willful discrimination by proxy. In such cases, a company using algorithmic differentiation does not aim to discriminate against a certain group but does so by accident. For instance, an electricity company might find that, during the previous five years, 5% of the people in Proxy group 1 run into serious payment troubles within five years. That 5% becomes unreachable while they still owe the electricity company money; on average €1,000. So, for 5% of Proxy group 1, the company wrote off an average of 1000 Euro during the past 5 years.

Say that the electricity company requires, from now on, all people from Proxy group 1 to pay a deposit of 1000 Euro if they enter an electricity contract. (Some electricity companies have such a policy.[28]) We assume that, in this case, the electricity company does not know that Proxy group 1 consists of predominantly people with a dark skin color.

For people in the same neighborhood that are not part of Proxy Group 1 (they drive other cars, or are born in a different period), the electricity company never had to write off debts during the past five years. For that group, the electricity company asks for a 100 Euro deposit. In that group, 95% of the people are white.

The result is that people with a dark skin color have a high chance that they have to pay a 1000 Euro deposit for an electricity contract. People with a white skin color almost never have to pay the 1000 Euro deposit. "It is possible," Custers et al. note, "that data mining will inadvertently use proxies for factors which society finds socially unacceptable, such as race, gender, nationality or religion."[29]

Hence, a company might inadvertently and even unconsciously discriminate against people with a certain skin color. In addition, the victims of this form of discrimination may not realize that they are being discriminated against.

We will see below that, from a legal perspective, such accidental discrimination can usually be characterized as indirect discrimination.[30] Roughly speaking, indirect discrimination occurs when a practice is neutral at first glance but ends up discriminating against people with a protected characteristic, such as a certain ethnicity or gender. Some legal protection against such indirect discrimination is thus possible.[31] The question to be addressed below, however, is whether current systems of European non-discrimination law are sufficiently able to deal with new types of unfair discrimination that have little to do with characteristics such as ethnicity.

---

28. Peter Kulche, *Onterecht een lage kredietscore én geschoffeerd [Wrongly being given a low credit score and being insulted]*, CONSUMENTENBOND [DUTCH CONSUMERS ORGANISATION] (Dec. 22, 2016), https://www.consumentenbond.nl/internet-privacy/onterecht-een-lage-kredietscore-en-geschoffeerd.
29. Custers, *supra* note 27.
30. *See infra* Part 3.

### C. *Algorithmic Decision-making and Harm to Non-protected Groups*

Algorithms can differentiate on the basis of non-protected grounds, which do not correlate with protected characteristics such as ethnicity or gender. Algorithms can find complicated correlations, whereby many factors taken together help to predict something about a group or an individual. Even though this might concern characteristics that are not included in traditional lists of protected grounds, such as ethnicity or gender, we suggest that such differentiation could still be unfair. As Custers et al. note, "discrimination might be transferred to new forms of population segments, dispersed throughout society and only connected by some attributes they have in common."[32]

For example, suppose that an electricity company finds a correlation between payment troubles and the following characteristics: people who drive a car of the type X of the brand Y, and live in postal code ABCD, and are born between 1981 and 1987. As in the earlier example, in the past five years, the company had to write off, on average, €1,000 of debt for 5% of the people in the group, because they left without a trace. We call this "Group 1."

There is a difference with our earlier example of Proxy Group 1. In Group 1, there is *no* correlation between the characteristics (car type etc.) and any protected characteristics, such as ethnicity. For instance, in Group 2 the percentage of each ethnicity is the same as in the country as a whole. Likewise, the age composition and the percentage of women in Group 2 is exactly the same as in the country as a whole. From the perspective of protected characteristics, Group 2 is completely normal and average, compared to the country as a whole. Hence, if somebody has characteristics that put him or her in Group 2 (owns that type of car, is born in a certain period, etc.), that does not predict anything about his or her ethnicity, gender, etc. (Admittedly, the hypothetical is far-fetched. We only present it for analytical purposes.)

---

31. *See* Council Directive 2000/43, art. 2(2)(b), 2000 O.J. (L 180) 22, 24 (EC); Council Directive 2000/78, art. 2(2)(b), 2000 O.J. (L 303) 16, 18; Council Directive 2004/113, art. 2(2)(b), 2004 O.J. (L 373) 37, 40; Directive 2006/54, art. 2(1)(b), 2006 O.J. (L 204) 23, 26.

32. Custers, *supra* note 27; *see also* Anton Vedder, *KDD: The challenge to individualism*, ETHICS & INFO. TECH. 1, 275–81 (1999).

Now suppose that the electricity company asks (as in the earlier example about Proxy Group 1) a deposit of €1,000 if people from Group 2 want to enter an electricity contract. In this case, the deposit requirement does not harm people with protected characteristics. After all, being a member of Group 2 does not correlate with protected characteristics. Nevertheless, even if the €1,000 deposit requirement did not harm people with a certain ethnicity or another protected characteristic, the practice could still be unfair, or at least controversial. We give three examples of such possible unfairness: such differentiation can (a) reinforce structural inequalities, (b) lead to decisions about people on the basis of seemingly irrelevant criteria, and (c) lead to incorrect predictions about individuals. We elaborate on these reasons in the following sections.

Our goal here is not to convince the reader that each form of algorithmic unequal treatment is unfair; unfairness depends on all the circumstances of a particular situation. We merely want to illustrate that algorithmic differentiation *can* lead to practices that are unfair, or at least controversial, even if the differentiation does not necessarily harm people with protected characteristics.

### 1. Reinforcing Structural Inequalities

One risk with algorithmic differentiation is that it could reinforce structural inequalities in society, in particular inequalities between socio-economic groups.[33] For instance, online stores could use algorithms to adapt prices to individual consumers: online price differentiation. An online store could charge different prices to different consumers, for identical products, based on information the store has about those consumers. There are examples where such pricing schemes have led to higher prices for people in low-income groups.

To illustrate: an online store in the U.S. charged more to people in rural areas than to people living in large cities. In large cities, there tend to be many stores. So, if consumers in a large city find the prices in an online store too high, they can easily visit a competing (brick and mortar) store. Therefore, the online store charged low prices in big cities, to ensure that consumers would not go to competing stores. In rural areas, someone who finds the online store too expensive may have to travel for hours to visit a competing (brick and mortar) shop. The online store therefore charged relatively high prices to customers in rural areas. In the United States, people in large cities are, on average, richer than people in the countryside. The price differentiation therefore led, probably unintentionally, to richer people paying lower prices.[34] Hence, the price differentiation reinforced financial inequality.

---

33. *See* Christian Neuhauser, *Relative Poverty as a Threat to Human Dignity: On the Structural Injustice of Welfare States*, 14 STUD. GLOB. JUST. ETHICAL ISSUES POVERTY ALLEVIATION 151, 161 (Helmut P. Gaisbauer et al. eds., Springer 2016) [hereinafter Christian Neuhauser].

In our hypothetical example, the electricity company requires a higher deposit from people in Group 2, because historical data predicts that the company has to write off debt for 5% of the people in that group. Let's assume that the majority of the people in Group 2 are relatively poor. In that case, people who are already struggling financially, have to pay a deposit of €1,000. People who are not part of Group 2 have to pay a deposit of only €100. Let's assume that these people (outside Group 2) are relatively wealthy. All in all, the result is that poorer people must pay a higher deposit.

In addition, a bank that lends money to people in Group 2 might charge higher interest rates, in response to the higher chance of people defaulting on a loan. In that case, poor people pay a higher interest rate: another example of reinforcing financial inequality.[35]

In sum, at least in some cases, algorithmic differentiation could lead to a situation where "the poor pay more".[36] Scholars such as Eubanks warn that algorithms could have many effects that "punish the poor."[37] Consequently, algorithmic decision-making might deepen existing socio-economic dividing lines and possibly create new disadvantages for groups that already find themselves in a socially vulnerable position.38 This, in turn, might affect their well-being and even human dignity in hitherto unknown ways.39

---

34. *See* Jennifer Valentino-DeVries et al., *Websites Vary Prices, Deals Based on Users' Information*, WALL ST. J. (Dec. 24, 2012), http://online.wsj.com/article/SB10001424127887323777204578189391813881534.html; *see also* Desai & Kroll, *supra* note 6, at 1.

35. We leave aside, for now, the question whether the bank's behavior is socially desirable. We merely want to illustrate that algorithmic price differentiation can reinforce financial inequality.

36. Phrase borrowed from David Caplovitz, *The Poor Pay More: Consumer Practices of Low-Income Families*, FREE PRESS (1963).

37. *See generally* VIRGINIA EUBANKS, AUTOMATING INEQUALITY: HOW HIGH-TECH TOOLS PROFILE, POLICE AND PUNISH THE POOR (Martin's Press 2018); *see also* Zarsky, *supra* note 13, at 124–27.

### 2. Seemingly Irrelevant Criteria

Another controversial aspect of algorithmic differentiation concerns seemingly irrelevant criteria. In our Group 2 hypothetical, an electricity company makes decisions about people, based on the fact that people from Group 2 have certain characteristics, namely they drive a car of the type X of the brand Y, live in postal code area ABCD, and are born between 1981 and 1987. One company makes decisions about the required deposit (€1,000 for an electricity contract); another company charges those people a higher interest rate for obtaining credit.

The people in Group 2 might feel like they are being punished or at least unfairly treated. Consumers might find it reasonable if a store checks her credit rate and salary before granting credit. But for a consumer it may feel unfair if she is denied credit on the basis of seemingly irrelevant characteristics such as her postal code or the type of car she drives (at least, irrelevant from the consumer's perspective).

We can turn here also to the real-life example that we mentioned in the introduction to this paper. In the Netherlands, some insurance companies charged extra for car insurance if customers lived in apartments with certain types of numbers (a letter included in the number, such as 4A or 20C).40 Presumably, the companies found a positive correlation (in historical data) between living at an address with such a house number and the chance of being in a car accident. Insurance companies can use such correlations to predict accidents, even if this correlation does not imply causation. Causation is not that important for the insurer, as long as the house number sufficiently reliably predicts accidents.

However, many people might think that it is strange if they mustpay more for their car insurance simply because they live at a certain house number. After all, what does one's house number have to do with driving qualities? People may think that it is unfair if they are punished for something that seems to be irrelevant for the case at hand.

---

38. *See* Krista Nadakavukaren Schefer, *The Ultimate Social (or is it Economic?) Vulnerability: Poverty in European Law*, *in* PROTECTING VULNERABLE GRPS.: THE EUR. HUM. RTS. FRAMEWORK 401 (Francesca Ippolito & Sara Iglesias Sánchez eds., Hart 2015) (identifying low-income groups as 'socially vulnerable groups') [hereinafter Krista Nadakavukaren Schefer]; *see also* Andrea Filetti & Jan G. Janmaat, *Income Inequality and Economic Downturn in Europe: A Multilevel Analysis of Their Consequences for Political Participation*, 53 ACTA POLITICA 327 (2018) (In turn socio-economic disparities may negatively impact on the political participation of low-income groups).

39. *See* Christian Neuh. . .user, *supra* note 33; Krista Nadakavukaren Schefer, *supra* note 38; *see also* Laurens Lavrysen, *Strengthening the Protection of Human Rights of Persons Living in Poverty Under the ECHR*, 33 NETH. Q. HUM. RTS. 293 (2015).

40. Gerald Koon, *supra* note 1.

The topic of predictions based on correlations, or seemingly irrelevant criteria, raises complicated questions. For now, our point is merely that people may find AI decisions unfair if an algorithmic system attaches consequences to a person's characteristics and those characteristics seem irrelevant. Such a feeling of unfairness has nothing to do with discriminating on the basis of ethnicity or other protected characteristics.

### 3. Incorrect Predictions

Another controversial aspect of algorithmic differentiation is that it can lead to incorrect predictions about individuals. The electricity company in our hypothetical uses a predictive model that says: for people in Group 2 (people who drive a car of the type X of the brand Y, live in postal code area ABCD, and are born between 1981 and 1987) there is a 5% chance that they will leave the company with an irrecoverable debt of, on average, €1,000. Let's assume in this case that the historical data, in aggregate, correctly predict future payments. The prediction does not say, however, whether a certain *individual* in Group 2 will default on payments. 95 % of the people in Group 2 will never miss a payment. Say, for example, that Alice has all the characteristics of Group 2, but she has never missed a payment in her life (she is part of the 95 %). For Alice, it may feel unfair if she has to pay a large deposit, simply because she forms part of a larger group that is considered to pose a greater risk for non-payment.[41]

In sum, we distinguished algorithmic differentiation (i) that directly or indirectly harms people with protected characteristics (such as people with a certain ethnicity), and (ii) that can lead to a disadvantage for people on the basis of non-protected characteristics that do not correlate with protected grounds. We suggested that such new, algorithmic differentiation can be unfair, even if it does not necessarily harm people with protected characteristics. For instance, such differentiation could reinforce financial inequality, could be based on seemingly irrelevant criteria, or could lead to incorrect predictions about individuals.

---

41. For a similar discussion, *See* Zarsky, *supra* note 13.

### D. *A Caveat*

A caveat is in order. For analytical purposes, we distinguished between differentiation (i) that harms people with protected characteristics and that (ii) is unfair in another way. However, in practice, category (i) and (ii) cannot be neatly separated.

For instance, in many countries, a situation in which poor people pay higher prices (reinforcing financial inequality) also harms people with certain skin colors, because financial status often correlates with skin color.[42] For example, people with an immigrant background are often poorer than people who do not have such a background.

Another complication concerns incorrect predictions. Above, we presented incorrect predictions as a type of harm that is different from the harm done by unequal treatment based on protected characteristics.[43] However, people who are part of a minority may suffer incorrect AI decisions more often than the majority.[44]

As Hardt notes, "it's true by definition that there is always proportionately less data available about minorities. This means that our models about minorities generally tend to be worse than those about the general population."[45] That problem arises when an algorithm is used to predict something which happens in a different frequency in the minority group. Hardt summarizes: "The lesson is that statistical patterns that apply to the majority might be invalid within a minority group."[46] In sum, a minority may suffer incorrect decisions more often than the majority. In such situations, the incorrect predictions problem could be, partly, classified under the header "harm to protected groups."

More generally, we therefore emphasize that there is no clearcut difference between (i) discrimination that harms people with protected characteristics and (ii) differentiation that is unfair in other ways. Although we make the distinction for analytical purposes in this paper, in real life, category (i) and (ii) may overlap.

---

42. On algorithms that reinforce social inequality, *see generally supra* Part 2.3.1.
43. *See supra* Part 2.3.3.
44. David Robinson et al., *Civil Rights, Big Data, and Our Algorithmic Future: A September 2014 Report on Social Justice and Technology*, UPTURN 12, 12–13 (2014) [hereinafter Robinson Report], https://bigdata.fairness.io/wp-content/uploads/2015/04/2015-04-20-Civil-Rights-Big-Data-and-Our-Algorithmic-Future-v1.2.pdf; *see also* Moritz Hardt, *How Big Data is Unfair. Understanding Unintended Sources of Unfairness In Data Driven Decision Making*, MEDIUM (Sept. 26, 2014), https://medium.com/@mrtz/how-big-data-is-unfair-9aa544d739de [hereinafter Hardt].
45. Hardt, *supra* note 44.
46. *Id.*

Now we turn to law, starting with a short introduction to non-discrimination law.

## II.  EUROPEAN NON-DISCRIMINATION LAW – A BRIEF INTRODUCTION

This paper has as its main aim to see which system of non-discrimination law is best suited in the European context to deal with the characteristics of algorithmic discrimination as we have described them in Section I. For readers who are not familiar with European non-discrimination law, this Section briefly introduces the main systems of non-discrimination law as they are currently in place as part of the European Convention on Human Rights (Section II(A)) and in the European Union (Section II(B)).[47] Most European countries also include a right to non-discrimination in their constitutions and have national non-discrimination statutes. However, as noted in the introduction, a further analysis of the various national non-discrimination law systems is outside the scope of this paper.

### A.  *The European Convention on Human Rights*

The European Convention on Human Rights (hereinafter: "ECHR") is a core human rights treaty that has been adopted in 1950 in the context of the Council of Europe.[48] The Council of Europe has 47 member states and is the most important international human rights organization in Europe. Compliance with the European Convention on Human Rights is supervised by the European Court of Human Rights (hereinafter: "Court").

The Court may receive applications from any person claiming to be the victim of a violation of her rights by one of the States Parties to the Convention.[49] The Court can decide that a State has violated the rights of the applicant, and the judgments are binding for the respondent states.[50] In addition, the general principles of interpretation as laid down in the Court's caselaw have so-called *res interpretata*, which means that states are bound to comply with them, even if they are laid down in judgments against other states.

---

47. *See generally* EUR. FUNDAMENTAL RTS. AGENCY & EUR. CT. H.R., HANDBOOK ON EUROPEAN NON-DISCRIMINATION LAW (2018), https://www.echr.coe.int/Documents/Handbook_non_discri_law_ENG.pdf [https://perma.cc/FN7G-CQTL] [hereinafter EUR. FUNDAMENTAL RTS. AGENCY & EUR. CT. HUM. RTS.].
48. Convention for the Protection of Human Rights and Fundamental Freedoms, Y.B. EUR. CONV. H.R. (2021) [hereinafter ECHR].
49. *Id.* art. 34.
50. *Id.* art. 46.

Accordingly, the Court can play an important role in developing human rights standards and principles.[51]

An applicant can also complain that a State did not sufficiently protect him or her against human rights violations by private actors, such as companies.[52] The Court can thus, indirectly, apply human rights to the relation between private actors.[53] In that way, the Convention rights have some "horizontal effect", that is, they have an influence on the relation between private actors.[54]

The Convention may play a role in cases on discrimination because of Article 14 ECHR, which reads as follows:

"The enjoyment of the rights and freedoms set forth in this Convention shall be secured without discrimination on any ground such as sex, race, colour, language, religion, political or other opinion, national or social origin, association with a national minority, property, birth or other status."[55]

Based on this provision, the Court has developed an extensive caselaw on the right to non-discrimination.[56] In that caselaw, the Court has also identified and detailed several key non-discrimination concepts, some of which are particularly relevant to algorithmic discrimination. We further discuss the most relevant of these in Section III.

## B. The European Union

The European Union (hereinafter, "EU") is a separate international organization, counting 27 member states, that was originally created to stimulate European economic co-operation and free movement of economically important goods (services, capital, workers, goods) between the member states. Protectionist treatment of a member state's own workers, services etc. would cause important impediments to well-functioning internal market.[57] Mainly for that reason, from an early stage, the EU's legislative and judicial institutions started to adopt measures and rulings to remove discriminatory barriers to free trade and free movement. Impediments could also be created by, for example, different labor conditions and unequal pay of male and female workers. For that reason, the EU has also adopted legislation to combat several forms of unequal treatment in the workplace.[58]

---

51. *See* Janneke H. Gerards, *The Paradox of the European Convention on Human Rights and the European Court of Human Rights' Transformative Power*, 4 KUTAFIN U. L. REV. 315 (2017).

52. *See* JANNEKE H. GERARDS, GENERAL PRINCIPLES OF THE EUROPEAN CONVENTION ON HUMAN RIGHTS 144 (Cambridge Univ. Press 2019).

53. *Id.*

54. *Id.*

55. *See* Protocol No. 12 to the Convention for the Protection of Human Rights and Fundamental Freedoms, art. 1, (Nov. 4, 2000) (laying down a similar prohibition, with, regarding certain aspects, a broader scope.). For an up-to-date list on signatories of protocol 12, *see Search on Treaties*, COUNCIL OF EUR. https://www.coe.int/en/web/conventions/search-on-treaties/-/conventions/treaty/177/signatures?p_auth=0Kq9rtcm
(last visited Sept. 21, 2021) (On Oct. 16, 2020, the total number of ratifications of accessions to Protocol 12 was twenty); *see also* ECHR, *supra* note 48, at 32.

56. *See* JANNEKE GERARDS, PROHIBITION OF DISCRIMINATION, IN PIETER VAN DIJK, FRIED VAN HOOF, ARJEN VAN RIJN, LEO ZWAAK (EDS), THEORY AND PRACTICE OF THE EUROPEAN CONVENTION ON HUMAN RIGHTS, 5TH EDITION, 2018 (showing an overview of the case law).

Over time, the original European Economic Community developed from a mainly economically oriented international co-operation of states to a more encompassing European Union, of which the main objectives nowadays include the protection of human rights and non-discrimination for their own sake.[59] A clause was added to the EU Treaties in 1997 to allow EU institutions – the Council of Ministers together with the European Parliament – to adopt non-discrimination legislation:

"[T]he Council, acting unanimously in accordance with a special legislative procedure and after obtaining the consent of the European Parliament, may take appropriate action to combat discrimination based on sex, racial or ethnic origin, religion or belief, disability, age or sexual orientation."[60]

Based on competence clauses such as this one, the EU institutions have adopted several non-discrimination laws. Mostly, these come in the form of Directives, which must be transposed into national law by the member states.[61] The four most important non-discrimination Directives, that are central to the present study, are the following:[62]

---

57. *See* MARK BELL, THE PRINCIPLE OF EQUAL TREATMENT: WIDENING AND DEEPENING 611 (Paul Craig & Gráinne de Búrca eds., 2011); *see* Janneke H. Gerards, *Non-Discrimination, the European Court of Justice and the European Court of Human Rights: Who Takes the Lead?*, *in* Part II THE EUROPEAN UNION AS PROTECTOR AND PROMOTER OF EQUALITY 135 (Thomas Giegerich ed., 2020).

58. *See* CATHERINE BARNARD, SEX EQUALITY LAW IN THE EUROPEAN UNION 321 (Tamara K. Hervey & David O'Keeffe eds., Wiley 1996).

59. Instead of as mere instruments to advance free trade and free movement. *See* Consolidated Version of the Treaty on European Union, 2012 O.J. (C 326) 13, arts. 2, 3, & 10, [hereinafter TEU].

60. Consolidated Version of the Treaty on the Functioning of the European Union, 2008 O.J. (C 115) 47, art. 19(1) [hereinafter TFEU] (formerly Article 13 of the Treaty on the European Communities, added to it in the Treaty of Amsterdam); *see* Treaty of Amsterdam Amending the Treaty on European Union, the Treaties Establishing the European Communities and Certain Related Acts, 1997 O.J. (C 340) 1, art. 13.

61. *See* TFEU, *supra* note 60, art. 288; *see* EUR. FUNDAMENTAL RTS. AGENCY & EUR. CT. HUM. RTS., *supra* note 47, at 21.

62. *See also* EUR. FUNDAMENTAL RTS. AGENCY & EUR. CT. HUM. RTS., *supra* note 47, at 22.

The Racial Equality Directive (2000) prohibits discrimination on the basis or racial or ethnic origin in many contexts.[63]

The Employment Equality Directive (2000) prohibits discrimination on the grounds of religion or belief, disability, age, or sexual orientation in the employment context.[64]

The Gender Goods and Services Directive (2004) prohibits discrimination on the basis of gender in the context of the supply of goods and services.[65]

The so-called Recast Directive (2006) prohibits discrimination based on gender in employment and occupation.[66]

As can be seen from these short descriptions, these Directives also apply to companies as long as their acts come within the material scope of application of the Directives. Therefore, both private actors and national authorities have an obligation to apply the provisions of these Directives as they are transposed in national law.

To ensure uniform and correct interpretation and application of these Directives in all EU member states, the Treaty on the Functioning of the European Union provides for a so-called preliminary reference procedure.[67] If one of the national courts is confronted with a question regarding the validity or interpretation of a provision of EU law, the court usually has to refer that question to the Court of Justice of the EU (hereinafter: "CJEU") to have it answered and have the meaning of a particular concept or phrase in the Directives authoritatively explained.[68] This procedure has given the CJEU significant opportunity to elucidate and develop some core non-discrimination notions such as direct and 'ndirect discrimination. These notions are further addressed in Section III.

In addition to the various non-discrimination Directives, the EU has its own human rights document, the Charter of Fundamental Rights of the European Union (2000) (hereinafter, "EU Charter"), which includes the rights to equality and non-discrimination.[69] Since 2009, this document is binding on both the institutions of the EU and on the authorities of the Member States insofar as they implement EU law.[70]

---

63. Council Directive 2000/43, 2000 O.J. (L 180) 22 (EC).
64. Council Directive 2000/78, 2000 O.J. (L 303) 16 (EC).
65. Council Directive 2004/113, 2004 O.J. (L 373) 37 (EU).
66. Council Directive 2006/54, 2006 O.J. (L 204) 23 (EU).
67. TFEU, *supra* note 60, art. 267.
68. *See* Srl Cilft v. Ministry of Health, (1982) E.C.R. 03415 (Eur.).
69. Charter of Fundamental Rights of the European Union, 2010 O.J. (C 83) 02, arts. 20–21 [hereinafter CFREU]; *see* EUR. FUNDAMENTAL RTS. AGENCY & EUR. CT. HUM. RTS., *supra* note 47, at 22.
70. TEU, *supra* note 59, art. 6(1); CFREU, *supra* note 69, art. 51(1); *see also* EUR. FUNDAMENTAL RTS. AGENCY & EUR. CT. HUM. RTS., *supra* note 47, at 23.

### C. EU Data Protection Law and its Relation to Non-Discrimination Law

Finally, the EU has become famous for its legislation on the protection of personal data since the General Data Protection Regulation (hereinafter: "GDPR") entered into force in 2018.[71] As a Regulation the GDPR is a different type of legislative instrument from the Directives described above. A Regulation does not need to be transposed into national law but must be applied directly on the national level.[72] The GDPR further has been designed so as to be binding on both public authorities and private actors, such as companies.[73]

Although the GDPR focuses on the (guarantees on fairness and transparency of) processing of personal data, several provisions can be relevant to issues of discrimination.[74] For instance, in principle, the GDPR contains a prohibition (subject to exceptions) of using of certain types of sensitive data, such as data about ethnicity:

> Processing of personal data revealing racial or ethnic origin, political opinions, religious or philosophical beliefs, or trade union membership, and the processing of genetic data, biometric data for the purpose of uniquely identifying a natural person, data concerning health or data concerning a natural person's sex life or sexual orientation shall be prohibited.[75]

This prohibition on the processing of personal data like ethnic origin or sexual orientation can help to prevent those public authorities and private actors make decisions that are based on such characteristics.[76] Hence, the GDPR partly aims for, and can contribute to, the protection against discrimination.[77]

---

71. Commission Regulation 2016/679, 2016 O.J. (L 119) 1 [hereinafter GDPR].
72. TFEU, *supra* note 60, art. 288.
73. Both public authorities and private actors can be 'processors' of personal data. *See* GDPR, *supra* note 71, art. 4(8).
74. *See* Philipp Hacker, *Teaching Fairness to Artificial Intelligence: Existing and Novel Strategies Against Algorithmic Discrimination Under EU Law*, 55 COMMON MKT. L. REV. 1143, 1171 (2018); Zuiderveen Borgesius, *supra* note 16, at 1578–82 [hereinafter Borgesius].
75. GDPR, *supra* note 71, art. 9(1).
76. However, the ban on using "special categories of personal data" also create challenges for mitigating discrimination. Many of the methods to tackle discrimination in AI systems implicitly assume that organizations hold these special category data. But many organizations may not hold such data, because of the GDPR's ban. See Marvin van Bekkum & Frederik Zuiderveen Borgesius, *Using special category data to prevent discrimination: does the GDPR need a new exception?* (forthcoming).
77. *See also* GDPR, *supra* note 71, pmbl. 71 ("In order to ensure fair and transparent processing in respect of the data subject . . . the [data] controller should . . . prevent, inter alia, discriminatory effects on natural persons on the basis of racial or ethnic origin,

In relation to algorithmic differentiation, the GDPR's specific rules on profiling and fully automated decision-making can also be relevant. Article 22(1) says:

"The data subject shall have the right not to be subject to a decision based solely on automated processing, including profiling, which produces legal effects concerning him or her or similarly significantly affects him or her."[78]

Thus, roughly summarized, it is not allowed to use fully automated systems to generate decisions about individuals with legal or similar effects. The GDPR also has specific transparency requirements for such fully automated decisions. However, so far, the provisions on automated decision-making (and their predecessors) have hardly been applied in practice.[79] Only the future can tell whether these provisions will have much practical effect.

As can be gleaned from the above, the GDPR is an important complement to the EU non-discrimination Directives, and the interrelationship between the GDPR and EU non-discrimination laws deserves further study to see how the interaction between these instruments can help protect against algorithmic discrimination.[80] Nonetheless, for the purposes of the present study we leave the GDPR out of consideration. The objective of our study is not, after all, to examine whether, how and to what degree data protection law can help to protect against algorithmic discrimination, but what system of non-discrimination law is best suited to deal with the specific challenges related to this type of discrimination. An exclusive focus on non-discrimination law fits this purpose best.

## III. OPEN, CLOSED, AND HYBRID SYSTEMS OF NON-DISCRIMINATION LAW

In Section I, we identified two distinct yet related issues regarding algorithmic decision-making: (i) decisions can lead to harm

---

political opinion, religion or beliefs, trade union membership, genetic or health status or sexual orientation, or [processing] that result[s] in measures having such an effect.").

78. GDPR, *supra* note 71, art. 22(1).

79. One exception in the application of the GDPR's provisions on automated decision-making in which a judge required a company to provide transparency about an automated decision. *See* Raphael Gellert, et. al., *The Ola & Uber judgments: for the first time a court recognises a GDPR right to an explanation for algorithmic decision-making*, EU L. ANALYSIS (Apr. 28, 2021, 7:06 AM), http://eulawanalysis.blogspot.com/2021/04/the-ola-uber-judgments-for-first-time.html [https://perma.cc/LVJ3-Q74E].

80. *See* Hacker, *supra* note 74, at 1172 (for preliminary endeavors in this regard); Janneke Gerards & Raphaële Xenidis, *Special Report on the Algorithmic Discrimination In Europe, European Equality Law Network*, EUR. COMM'N (Mar. 10, 2021), https://op.europa.eu/s/pkNW [https://perma.cc/LJ57-9HK3]; Borgesius, *supra* note 74, at 1572.

to people with protected characteristics, such as ethnicity or gender; (ii) algorithmic decision-making can lead to other forms of unfair differentiation and structural inequalities, which have little to nothing to do with characteristics that receive explicit protection in non-discrimination law.

As mentioned above, these two issues raise a question. Are existing nondiscrimination laws, such as those briefly discussed in Section II, capable of addressing discrimination-related risks of algorithmic decision-making? To answer this question, we distinguish three theoretically conceivable systems of nondiscrimination and equal treatment law: (i) fully open systems, (ii) fully closed systems, and (iii) hybrid systems.[81] In the current section, we further explore these main systems and their subtypes. In addition, we briefly set out their interrelationship. To illustrate their practical relevance, we will also show how and to what extent these theoretical possibilities are reflected in the European nondiscrimination laws that we mentioned in Section II.

## A. *Fully Open Systems*

The first theoretically possible system is a fully open system, both regarding the protected grounds and as regards possible exemptions. To give an example of such a clause: Article 20 of the EU Charter, discussed in Section II(A), states that "everyone is equal before the law."[82] This clause is fully open in that it contains no exhaustive list of protected grounds. The provision does not even mention any particular grounds of discrimination such as ethnicity, skin color, or gender. Open nondiscrimination provisions therefore may be considered (in theory) to cover inequalities based on the type of one's car just as much as discrimination based on one's sexual orientation or ethnicity. Second, Article 20 EU Charter is fully open as to the situations in which exemptions can be made to the prohibition of discrimination, or, put differently, as to the situations in which there is no problematic unequal treatment. Hence, courts or bodies applying an open provision such as this one are free to identify any differences between persons, groups, or situations that they consider relevant or any other reasons or considerations that may justify a difference in treatment.

---

81. *See* Aalt Willem Heringa, *Standards of Review for Discrimination: The Scope of Review by the Courts*, NON-DISCRIMINATION LAW: COMPARATIVE PERSPECTIVE 25 (Titia Loenen & Peter R. Rodrigues eds., Kluwer Law Int'l 1999) [hereinafter Heringa]; Janneke H. Gerards, *Gronden van discriminatie - de wenselijkheid van open en gesloten opsommingen [Grounds of Discrimination - The Desirability of Open and Closed Lists of Grounds]*, DE NIEUWE FEDERALE ANTIDISCRIMINATIEWETTEN LES NOUVELLES LOIS LUTTANT CONTRE LA DISCRIMINATION [THE NEW FEDERAL ANTI-DISCRIMINATION LAWS], 129 (Christian Bayart et al. eds., die Keure la Charte 2008); Janneke H. Gerards, *Nieuwe ronde, nieuwe kansen: naar een 'semi-open' systeem van gelijkebehandelingswetgeving? [New Round, New Chances: Toward a 'Semi-Open' System of Equal Treatment Law?]*, 36 NEDERLANDS TIJDSCHRIFT VOOR DE MENSENRECHTEN [NETH. J. HUM. RTS.] 144 (2011).

82. CFREU, *supra* note 69, art. 20.

### B. Fully Closed Systems

Fully closed systems are systems which both contain an exhaustive list of prohibited grounds of discrimination and a limited number of exemptions allowing for unequal treatment on such grounds.[83] Although such closed systems can exist in theory, they are less often seen in practice. Nevertheless, a closed system can be seen to exist in most of the EU Directives prohibiting so-called direct discrimination.[84] For example, the Employment Equality Directive (2000/78) exhaustively lists four grounds on which discrimination in employment is prohibited: religion or belief, disability, age, and sexual orientation.[85] Similarly, the Racial Equality Directive 2000/43 prohibits direct discrimination only on racial or ethnic origin,[86] and the Goods and Services Directive (2004/113) and the Recast Gender Equality Directive (2006/54) specifically prohibit unequal treatment of "men and women."[87]

The EU Directives thus concentrate on a limited, closed list of grounds. Once it is demonstrated that a difference in treatment is based directly, exclusively, or decisively on one of these grounds, the prohibition applies. The prohibition of direct discrimination does not apply if the difference in treatment is, for example, based on a ground such as postal code or a foreign sounding last name.

There are One can imagine various exceptions to the above-mentioned bans on direct discrimination, but the system is fully closed. [88] This means that all Directives contain an exhaustive list of situations and conditions for accepting certain forms of unequal treatment on one of the covered grounds.[89] For example, the Racial

---

83. Heringa, *supra* note 81, at 27.
84. For a brief description of the Directives and further sources, see supra, Section 3. *See also* Heringa, *supra* note 81;
EUR. FUNDAMENTAL RTS. AGENCY & EUR. CT. HUM. RTS., *supra* note 47, at 96–108, and at 155–228.
85. *See* Council Directive 2000/78, at 11, 2000 O.J. (L 303) 16 (EC).
86. Council Directive 2000/43, at art. 1, 2000 O.J. (L 180) 22 (EC).
87. *See* Council Directive 2004/113, at art. 1, 2004 O.J. (L 373) 37 (EU); *see* Council Directive 2006/54, at art. 1, 2006 O.J. (L 204) 23 (EU).
88. *See* Mark Bell, *Direct Discrimination*, CASES, MATERIALS & TEXT ON NAT., SUPRANATIONAL & INT. NON-DISCRIMINATION L. 185, 269 (Dagmar Schiek et al. eds., Hart 2007); EUR. FUNDAMENTAL RTS. AGENCY & EUR. CT. HUM. RTS., *supra* note 48, at 96–108.
89. *See also* EUR. FUNDAMENTAL RTS. AGENCY & EUR. CT. HUM. RTS., *supra* note 47, at 34.

Equality Directive only allows for differences in treatment based on racial or ethnicity if such a characteristic "constitutes a genuine and determining occupational requirement,"[90] or if it is necessary to adopt or maintain "specific measures to prevent or compensate for disadvantages linked to racial or ethnic origin."[91] All other instances of direct discrimination based on these grounds that fall within the material scope of application of the Directive are prohibited.

## *C. Hybrid Systems*

As mentioned, next to fully open and fully closed systems some in-between or hybrid options can be conceived of. Such hybrid systems can combine elements of fully open and fully closed systems in two respects: (i) where their list of grounds is concerned, and (ii) regarding the possibilities for justification of unequal treatment (or, put differently, the possibilities for exemption). We deal with hybrid systems as to the lists of grounds first and then turn to hybrid systems as regards the exemptions.

### 1. Hybrids as Regards Lists of Grounds

First, prohibitions of discrimination may have a "semi-closed" list of grounds. This means that they contain a list of grounds that, on first glance, seems to be closed, because it contains a clear enumeration of grounds, but a closer look at the text reveals that the list—in reality—is open.[92] For example, Article 21 of the EU Charter states:

"Any discrimination *based on any ground such as* sex, race, colour, ethnic or social origin, genetic features, language, religion or belief, political or any other opinion, membership of a national minority, property, birth, disability, age or sexual orientation shall be prohibited (emphasis added)."[93]

Although this provision contains a clear list of grounds, the list is not closed; the phrase "any ground such as" shows that the list is open-ended and therefore only "semi-closed." Such a semi-closed list can also be found in Article 14 ECHR:

"The enjoyment of the rights and freedoms set forth in this Convention shall be secured without discrimination *on any ground such as* sex, race, colour, language, religion, political or other opinion, national or social origin, association with a national minority, property, birth *or other status*."[94]

---

90. Council Directive 2000/43, 2000 O.J. (L 180) 22 (EC).
91. *Id.* at art. 5.
92. *See also* EUR. FUNDAMENTAL RTS. AGENCY & EUR. CT. HUM. RTS., *supra* note 47, at 161–62.
93. CFREU, *supra* note 70, art. 21.

Here, the semi-closed nature of the list is illustrated by the use of wording such as "any ground such as" and the addition of "other status" to the listed grounds.[95]

In such semi-closed lists, the question may arise as to the function of the enumerated grounds, and the actual difference between such semi-closed and fully open clauses? After all, from a legal perspective, non-exhaustive lists of grounds would seem to allow complaints to be made related to differentiation based on all kinds of grounds, ranging from one's postal code to one's preferences for certain foodstuffs.[96]

The difference between fully open lists of grounds and semi-closed lists is that the semi-closed nature of the list of grounds may bring along certain restrictions as to which grounds courts can add to the list. For example, the European Court of Human Rights has accepted that the (seemingly open) notion of other status must be read in line with the listed grounds, which serve as a kind of "example grounds".[97] Hence, the phrase other status does not cover all forms of differentiation.[98] This possibility of justification is more open and less clearly defined than the closed exemptions for direct discrimination.[99] We can therefore call these systems open from the perspective of exemptions as far as indirect discrimination is concerned. Consequently, the Directives contain a hybrid, semi-closed system of exemptions: They are fully closed as far as the justification of direct discrimination is concerned, but open as regards the justification of indirect discrimination.

In some cases, a semi-closed exemption may even exist in relation to prohibitions of direct discrimination.[100] An example is the system of exemptions for age-based discrimination provided in the Employment Equality Directive 2000/78. The Directive stipulates in general terms that both direct and indirect differences in treatment based on age do not constitute discrimination if they are "objectively and reasonably justified by a legitimate aim, including legitimate employment policy, labor market and vocational training objectives, and if the means of achieving that aim are appropriate and necessary."[101]

---

94. ECHR, *supra* note 48, art. 14 (emphasis added).
95. *See, e.g.*, FRÉDÉRIC EDEL, THE PROHIBITION OF DISCRIMINATION UNDER THE EUROPEAN CONVENTION ON HUMAN RIGHTS 86.
96. *See* Janneke H. Gerards, *The Discrimination Grounds of Article 14 of the European Convention on Human Rights*, 13 HUM. RTS. L. REV. 99, 104 (2013).
97. *Id.*
98. *Id.* at 104–05.
99. *See also* EUR. FUNDAMENTAL RTS. AGENCY & EUR. CT. HUM. RTS., *supra* note 47, at 103.
100. In nondiscrimination doctrine, this last hybrid exemption is usually typified as "semi-open" rather than "semi-closed" to reflect that, compared to the fully open possibility of justification that exists in relation to age-based discrimination, this exemption is more "closed." To avoid confusion, we have decided to not use that wording here and continue the "semi-closed" terminology instead.

This provision is reminiscent of the open system of exemptions for indirect discrimination, but it also applies to direct age discrimination.[102] The system is not fully open, however, since the provision continues to list a number of situations in which a justification may be considered to exist in any case, such as "the setting of special conditions on access to employment . . . for young people, older workers and persons with caring responsibilities . . . , the fixing of minimum conditions of age, professional experience or seniority . . . , the fixing of a maximum age for recruitment . . .", and "social security schemes of ages for admission or entitlement to retirement or invalidity pensions."[103] Thus, even though this system creates some clarity as to the potentially acceptable limitations to the equality principle, it allows for considerable leeway in practice.[104]

### D. Overview

In theory, it is possible to distinguish a spectrum of nondiscrimination laws, which range from fully open to fully closed systems, with possible hybrid options in-between. Moreover, we have shown that many such systems can also be seen to be reflected in the practice of European nondiscrimination law. The different systems and their manifestations can be summarized as follows:

| Type of system | Subtype of system | Example |
| --- | --- | --- |
| 1. Fully open | | Article 20 CFR |
| 2. Fully closed | | Most EU Directives insofar as direct discrimination is concerned |
| 3. Hybrid | 3a. Semi-closed list of grounds and fully open possibility of exemptions | Article 14 ECHR Article 21 CFR |

---

101. Council Directive 2000/78, art. 6, 2000 O.J. (L 303) 16, 19–20 (EC).
102. *See* EUR. FUNDAMENTAL RTS. AGENCY & EUR. CT. HUM. RTS., *supra* note 47, at 103–04.
103. Council Directive 2000/78, *supra* note 101, at 20.
104. *See also* EUR. FUNDAMENTAL RTS. AGENCY & EUR. CT. HUM. RTS., *supra* note 47, at 104.

|  |  |  |
|---|---|---|
|  | 3b. Closed list of grounds and fully open possibility of exemptions | Most EU non-discrimination Directives insofar as indirect discrimination is concerned |
|  | 3c. Closed lists of grounds and semi-closed list of exemptions | EU Directive 2000/78 insofar as the ground of age is concerned |

This table and the previous sections show that fully closed systems, which combine an exhaustive list of protected grounds and a closed set of possible exemptions for unequal treatment on these grounds, hardly exist in Europe. Even the EU nondiscrimination Directives, which come closest to a closed system, are expressive of a hybrid system since they allow for an open possibility of justification for indirect forms of discrimination.

In fact, the most relevant contrasts are between nondiscrimination legislation containing either semi-closed or fully closed lists of grounds (exemplified by Articles 14 ECHR and Article 21 EU Charter versus the EU Directives), and between nondiscrimination legislation containing open or semi-closed possibilities for exemptions (again exemplified by Articles 14 ECHR and 21 EU Charter versus the EU Directives). As noted, the EU Directives contain a special hybrid, since their system of exemptions for direct discrimination on one of the exhaustively listed grounds of discrimination is fully closed, whereas the system of exemptions for indirect discrimination on one of the protected grounds is fully open.

Thus, there appears to exist a gap between the theoretical spectrum of nondiscrimination systems that are available, and the use that is made of this in the reality of European nondiscrimination law. Nevertheless, in identifying which system would best be able to address the discrimination challenges set by AI, it is useful to explore the full array of theoretical possibilities, and not to narrow down the discussion to what is most common in European law. In Sections V and VI, this paper therefore sets out to identify the different qualities of all conceivable systems as well as explore which different systems could best be used in relation to the challenges set by AI-driven discrimination.

### E. Complication: Differences Between Grounds

Before identifying the specific characteristics, strengths and weaknesses of the different systems, a caveat must be added. In practice, provisions containing fully open or semi-closed lists of grounds (that is, subsystems 1 and 3a in the table provided in Section III(D) are not always applied in exactly the same way to discrimination based on different grounds. This difference in application has to do with the notion of "suspectness" of certain grounds of discrimination.[105]

To explain, in nondiscrimination law it is generally accepted that differences in treatment that are disadvantageous to particular people or groups should be based on an objective, neutral or reasonable justification. For a variety of reasons, scholars, legislators, and courts assume that it is difficult to provide such an objective and reasonable justification for distinctions that are made on grounds such as skin color or ethnicity.[106] Such grounds are 'suspected' to constitute an irrational and inherently unfair basis for decision-making.

Based on legal theory, four reasons for such "suspectness" can be given.[107] Put briefly, first, certain personal characteristics are immutable, and it can be considered unfair to put someone at a disadvantage because of characteristics that person cannot help having. Second, differentiation based on these grounds is usually informed by elements of undue stereotyping and prejudice related to certain characteristics (which means that usually there is no *objective* reason for distinguishing on the basis of these grounds). Third, groups characterized by such grounds are generally "vulnerable" or constitute "discrete and insular minorities" and therefore deserve special protection against disadvantageous treatment.[108] Fourth, there can be a history of social exclusion and stigmatization against groups characterized by certain personal traits, which may increase the likelihood that negative sentiments regarding these groups taint decision-making.

---

105. *See* JANNEKE H. GERARDS, JUDICIAL REVIEW IN EQUAL TREATMENT CASES 86 (2005); Janneke H. Gerards, *Intensity of Judicial Review in Equal Treatment Cases*, 51 NETH. INT'L L. REV. 135, 136 (2004) [hereinafter *Intensity*].

106. *See infra* Section 5.3 for more detail.

107. For further overviews of these reasons, *see*, *e.g.*, Andrew Altman, *Discrimination*, *in* THE STANFORD ENCYCLOPEDIA OF PHILOSOPHY (Edward N. Zalta ed. 2020), https://plato.stanford.edu/archives/sum2020/entries/discrimination [https://perma.cc/2E4X-Z272]; GERARDS, *supra* note 105.

108. *Compare* United States v. Carolene Prod.'s Co., 304 U.S. 144, 152 n.4 (1938) (the famous United States formulation of the "discrete and insular minority"), *with, e.g.*, Alajos Kiss v. Hungary, App. No. 38832/06, ¶ 42 (May 20, 2010), http://hudoc.echr.coe.int/eng?i=001-98800 [https://perma.cc/2VCU-EQNF] (providing an example of the more common use of the vulnerability criterion by the ECHR). *See generally* Alexandra Timmer, *A Quiet Revolution: Vulnerability in the European Court of Human Rights*, *in* VULNERABILITY: REFLECTIONS ON A NEW ETHICAL FOUNDATION FOR LAW AND POLITICS 147 (Martha A. Fineman & Anna Grear eds., Ashgate 2013) (on the concept of vulnerability).

In many systems containing closed or semi-closed lists of grounds, such 'suspect' grounds of discrimination are specifically mentioned. Almost all lists of grounds, including those in the ECHR and the CFR, refer to, for example, race, ethnicity and gender.[109] The lists thus reflect a societal rejection of certain grounds as permissible bases for decision-making.

Even if lists are fully open, however, the same type of consideration may inform how the justification for a difference in treatment is applied. For example, usually courts strictly assess the reasons advanced to justify a difference in treatment based on one of these suspect grounds, regardless of whether they are (exhaustively) listed.[110]

For instance, the ECHR has accepted that, generally, it will be problematic to base a difference in treatment on grounds such as ethnicity,[111] gender,[112] or sexual orientation.[113] This then results in strict scrutiny (which will often be "fatal in fact")[114] and require "very weighty" or "compelling" reasons for justification.[115] Thus, even in open systems that allow individuals to bring a case of unequal treatment on any conceivable ground before a court to be openly assessed for its fairness, the result of such strict review of distinctions based on suspect grounds is that they are more difficult to justify than distinctions based on more neutral grounds.[116] When choosing one or the other type of nondiscrimination system, this complicating factor should be considered.

## IV. CHARACTERISTICS, STRENGTHS, AND WEAKNESSES OF THE THREE NON-DISCRIMINATION SYSTEMS

In Section III we have shown that there is a theoretical spectrum to be distinguished in nondiscrimination laws, which runs from fully open to fully closed systems, with a number of hybrid systems in-between. For the purposes of finding an adequate legislative response to the challenges set by AI-driven discrimination, the three main systems of non-discrimination law all show particular characteristics, strengths, and weaknesses. In this part we therefore aim to identify characteristics of closed, open, and hybrid systems that should be considered when designing nondiscrimination laws. We pay attention to institutional issues, symbolic value, benchmarking and societal developments (Section IV(A); legal certainty, clarity, flexibility and rigidity (Section IV(B); the reasons for including certain grounds of discrimination (Section IV(C); and the difference between equality and nondiscrimination (Section IV(D).

---

109. CFREU, *supra* note 69, arts. 20–21, at 395–96; ECHR, *supra* note 48, art. 14.
110. *E.g.*, *Intensity*, *supra* note 106, at 162.
111. Timishev v. Russia, App. Nos. 55762/00 & 55974/00, ¶ 56 (Dec. 13, 2005), http://hudoc.echr.coe.int/eng?i=001-71627.
112. *See* L. & V. v. Austria, App. Nos. 39392/98 & 39829/98, ¶ 45 (Jan. 9, 2003), http://hudoc.echr.coe.int/eng?i=001-60876.
113. *Id.*
114. *See* Gerald Gunther, *Foreword: In Search of Evolving Doctrine on a Changing Court: A Model for a Newer Equal Protection*, 86 HARV. L. REV. 1, 8 (1972) (describing the 'strict in theory and fatal in fact' line of reasoning).
115. *See Intensity*, *supra* note 105, at 162.
116. *Id.* at 161.

### A. *Institutional Considerations, Symbolic Value, Benchmarking, and Societal Developments*

In drafting nondiscrimination laws, the choice for an open, closed, or hybrid system is often determined by institutional considerations. If a fully closed system is adopted, or a hybrid system with a semi-closed list of grounds and/or a semi-closed possibility for exemptions, it is usually the legislature who formulates the grounds and exemptions.[117] As we will further explain in Section IV(C), making such choices is a political rather than a purely technical or formal exercise. The legislative or constitutional inclusion (or exclusion) of certain grounds expresses the societal and political views regarding the suspect nature of certain grounds of discrimination.

For example, including the ground of sexual orientation in a closed or semi-closed list of grounds can be seen to be an express, political and collectively supported response to developments or situations that are considered to be unacceptable (e.g. hostilities towards same-sex partners or LGTBI-persons), and it may signal a widely shared desire to strive for emancipation and full societal inclusion.[118] The fact that such grounds are included in high-level legislation (perhaps even a constitution or an international treaty) therefore has strong symbolic value and it emphasizes societal rejection or acceptance of certain characteristics.[119]

Therefore, closed lists have the most symbolic value.[120] However, even in semi-closed lists, the expressly included grounds have a benchmark function, which means that they signal a rejection of certain characteristics as unacceptable bases for decision-making.[121] Such "example grounds" demonstrate which grounds of discrimination are clearly considered to be unacceptable.[122] Although new grounds can be added to a semi-closed list, the included grounds implicitly show that these new grounds should at least be similar in nature or suspectness to the listed grounds.[123] Thus, by adopting a semi-closed list, the legislature can still give clear indications to those who have to apply the non-discrimination clause as to which types of discrimination ought to be covered. Consequently, expressly mentioning certain grounds in prohibitions of discrimination has significant symbolic value.[124]

---

117. *See also* Heringa, *supra* note 81, at 28.
118. Gerards, *supra* note 81; Elisabeth Holzleithner, *Mainstreaming Equality: Dis/Entangling Grounds of Discrimination*, 14 TRANSNAT'L L. & CONTEMP. PROBS. 927, 953 (2004).
119. *See* Gerards, *supra* note 81; Holzleithner, *supra* note 118, at 932.
120. *See, e.g.*, Titia Loenen, *Wijzigen of handhaven artikel 1 Grondwet: Bij twijfel niet inhalen [Amending or Retaining Article 1 of the Netherlands Constitution: In Case of Doubt, Do Not Pass]*, 41 NEDERLANDS TIJDSCHRIFT VOOR DE MENSENRECHTEN [NETH. J. HUM. RIGHTS], no. 3, at 319 (2016).

In addition, nondiscrimination legislation with closed or semi-closed lists invites regular societal and political debate on whether certain grounds need to be added.[125] Indeed, closed and semi-closed lists of grounds (with their example and benchmark function) may need regular updates.[126] After all, discrimination and perceptions of discrimination are narrowly connected with their technological, cultural, societal and political context. If the context changes so will ideas on discrimination.[127] Merely a century ago, for instance, it was unthinkable that same-sex partners would claim the same rights as different-sex partners, that children born out of wedlock could inherit property, or that women could be granted voting rights.[128] Over time, these views have changed completely; it is now fully accepted that sexual orientation and gender are suspect grounds of discrimination that can only rarely constitute a reasonable basis for decision-making.[129]

---

121. *See, e.g.*, *id.*
122. *See, e.g.*, *id.*
123. *See, e.g.*, *id.*
124. *See, e.g.*, Christine Jolls, *Antidiscrimination Law"s Effects on Implicit Bias* (Yale L. & Econ. Rsch. Paper No. 343, Yale L. Sch., Pub. L. Working Paper No. 148, 2005), https://papers.ssrn.com/sol3/papers.cfm?abstract_id=959228 [https://perma.cc/JX5B-6SE5] (last visited Sept. 26, 2021) (describing the importance of this symbolic value for protecting against discrimination).
125. *See* Gerards, *supra* note 81.
126. For some scholars this is a reason to plead against (semi-)closed systems; *see, e.g.*, Altman, *supra* note 107, § 5; Janneke H. Gerards, *Artikel 1 Grondwet – goede gronden voor wijziging?* [*Article 1 of the Netherlands Constitution – Good Grounds for Change?*], 41 NEDERLANDS TIDJSCHRIFT VOOR DE MENSENRECHTEN [NETH. J. HUM. RTS.], no. 3, at 304 (2016).
127. *See, e.g.*, Jakob Cornides, *Three Case Studies on 'Anti-Discrimination'*, 23 EUR. J. INT'L L. 517, 519 (2012).
128. *See, e.g.*, Stanley I. Benn & Richard S. Peters, *Justice and Equality*, *in* THE CONCEPT OF EQUALITY 54, 64–65 (William T. Blackstone ed., 1969) (discussing easy acceptance of 'biological' inequalities between men and women).
129. *See* Cornides, *supra* note 127, at 522.

It is likely that this process of change will be perpetual and that we will continue to recognize that certain groups are treated unfairly or certain characteristics should not matter to our decision-making.[130] For example, grounds such as socio-economic status may gradually come to be recognized as suspect grounds of discrimination.[131] Inclusion of such new grounds in closed or semi-closed lists is the result of a growing societal consensus being codified in legislation, based on democratic decision-making and deliberation.[132] This is significant from an institutional perspective, and it is a particular strength of closed and semi-closed systems that they can showcase the results of such societal and democratic decision-making processes.

By contrast, such symbolic, societal, and democratic processes are less clearly reflected in open systems of discrimination. Surely it has some symbolic value to state in general terms that everyone has a right to equal protection, but this says little on what that means in practice, especially if a benchmark (such as a set of example grounds) is missing to help determine what constitutes equal protection.

Moreover, fully open systems mainly rely on judges explaining in concrete cases if the principle of equal treatment has been disrespected.[133] As we explained above, an open system allows a court to identify certain grounds of discrimination as more problematic than others and set higher demands on their justification. Indeed, by doing so courts often pave the way for subsequent codification or inclusion of new grounds of discrimination. Incrementally and on a case-by-case basis, courts can deem certain types of discrimination on new grounds to be unjustifiable, and they can thus contribute to the building of a consensus regarding the "suspectness" of such grounds. Nevertheless, the symbolic value of judgments is weaker than that of an amendment of constitutional, treaties, or legislative provisions. Additionally, court-made nondiscrimination law lacks the democratic legitimacy of the lawmaking processes.[134]

---

130. In more recent years, for example, there is a debate on whether appearance or weight should perhaps be considered protected grounds of discrimination; *see generally, e.g.*, Deborah Carr & Michael A. Friedman, *Is Obesity Stigmatizing? Body Weight, Perceived Discrimination, and Psychological Well-Being in the United States*, 46 J. HEALTH AND SOC. BEHAV. 244 (2005); *see generally, e.g.*, Rebecca M. Puhl et al., *Perceptions of Weight Discrimination: Prevalence and Comparison to Race and Gender Discrimination in America*, 32 INT'L J. OBESITY 992 (2008); THE EUROPEAN UNION AS PROTECTOR AND PROMOTER OF EQUALITY (Thomas Giegerich ed., Springer 2020).

131. On 'suspect' grounds, see *infra*, Section 4.2.

132. An example of this can be found in Charter art. 21, which was adopted in 2000 and therefore represents a more recent consensus on suspect grounds of discrimination than ECHR art. 14 does. Comparing the semi-closed lists of grounds of the two provisions shows that 'ethnic origin' (next to 'race' and 'national origin'), 'belief' (next to 'religion'), 'sexual orientation', 'disability', 'age' and 'genetic features' have been added, probably to reflect the newly grown understanding that these would seem to be *a priori* unfair grounds for differentiation. Specifically on the ground of genetic features, *see, e.g.*, GENETIC DISCRIMINATION: TRANSATLANTIC PERSPECTIVES ON THE CASE FOR A EUROPEAN LEVEL LEGAL RESPONSE (Gerard Quinn et al. eds, 1st ed., Routledge 2016).

133. Heringa, *supra* note 81, at 28; Gerards, *supra* note 81, at 136.

## B. Legal Certainty, Clarity, and Precision

### 1. Legal Certainty

An advantage of fully or semi-closed systems is that they create considerable legal certainty.[135] For the outside world such systems make clear from the outset what types of discrimination are prohibited, in which situations exemptions can be made, and what balance is struck between the interest of nondiscrimination and conflicting interests, such as freedom of contract or the right of personal autonomy.[136] Such clarity is the more valuable if nondiscrimination legislation is applicable to private actors such as employers and companies offering services or distributing goods. For private actors in particular, it is important to be able to predict the legal consequences of their choices and the limitations of their freedom of contract and individual autonomy. Similarly, members of protected groups may benefit from the clarity such closed or semi-closed systems have to offer. The fact that their grounds are expressly included in a certain list of grounds makes it easier for them to claim a right to equal protection in concrete cases. Closed systems therefore offer much in terms of clarity and legal protection.

Admittedly, courts can offer some such legal certainty also in an open non-discrimination system. Courts can build a consistent caselaw by applying legal standards for justification and by explaining the consequences of deeming certain grounds to be "suspect." Such caselaw makes clear to government bodies and private actors which forms of unequal treatment are prohibited, and victims of discrimination can invoke the relevant precedent. Nevertheless, it can take a considerable amount of time for such standards to be developed, not in the least because judicial decision-making is dependent on the types of cases that parties bring before the courts. At least in theory, legislators are able to create certainty as to the applicable legal standards and criteria faster and sooner than courts and closed or semi-closed systems of exemptions may also be preferred for that reason.

---

134. Gerards, *supra* note 81, at 136.
135. Heringa, *supra* note 81, at 33; Gerards, *supra* note 81, at 138.
136. Gerards, *supra* note 81, at 139.

### 2. Precision and Application to Multiple and Intersectional Discrimination

In general, we can see that the more elaborate and detailed a closed list of grounds is, the more elaborate and detailed the system of exemptions will be. Different sets of exemptions and differently phrased clauses and standards are often attached to the different grounds of discrimination to do justice to their particularities. For example, in relation to the ground of disability, specific obligations to provide for reasonable accommodation may be required; for gender, specific attention may be needed for gender-based harassment or violence; and in relation to the ground of religion or belief, specific exemptions may be needed to allow organizations and companies based on a religious ethos to uphold their faith. Such differences between the grounds and exemptions may also find a translation into specific possibilities for monitoring and supervision, for differentiation in the rules of evidence, or for having access to a (semi-judicial) equal treatment body.

Surely, as we mentioned in Section IV(B)(1), detailed and differentiated systems of non-discrimination law offer advantages in terms of clarity and legal certainty, but they also bring along at least two difficulties. First, it will be essential to ensure a good match between the particularities connected to a certain ground of discrimination and the applicable system of exemptions and protection. If these run out of sync, either from the beginning or because of societal developments, the system may have unfair outcomes or cause difficulties in its practical application.

Second, precise and highly differentiated systems of exemptions are often criticized for their inability to deal with intersectional and multiple discrimination.[137] Multiple or cumulative discrimination occurs if a case of discrimination is based on different

---

137. A significant amount has been written on these notions. *See, e.g.*, Elaine W. Shoben, *Compound Discrimination: The Interaction of Race and Sex in Employment Discrimination*, 55 N.Y.U. L. REV. 793 (1981); Kimberle Crenshaw, *Mapping the Margins: Intersectionality, Identity Politics, and Violence Against Women of Color*, 43 STAN. L. REV. 1241 (1991); TIMO MAKKONEN, MULTIPLE, COMPUND AND INTERSECTIONAL DISCRIMINATION: BRINGING THE EXPERIENCES OF THE MOST MARGINALIZED TO THE FORE (Institute For Human Rights, Åbo Akademi University, 2002); Sarah Hannett, *Equality at the Intersections: The Legislative and Judicial Failure to Tackle Multiple Discrimination*, 23 OXFORD J. LEGAL. STUD. 65 (2003); Elisabeth Holzleithner, *supra* note 118; Sandra Fredman, *Double Trouble: Multiple Discrimination and EU Law*, EUR. ANTI-DISCRIMINATION L. REV., no. 2, Oct. 2005, at 13; *See also, e.g.*, EUR. FUNDAMENTAL RTS. AGENCY & EUR. CT. HUM. RTS., *supra* note 47, at 59; Catherine E. Harnois, *Jeopardy, Consciousness, and Multiple Discrimination: Intersecting Inequalities in Contemporary Western Europe*, 30 SOCIO. F. 971 (2015); Dagmar Schiek, *Intersectionality and the Notion of Disability in EU Discrimination Law*, 53 COMMON MKT. L. REV. 35 (2016); Altman, *supra* note 107, §7.

grounds at the same time, for example if an employer refuses to give a job to a Roma man who is the single caretaker for a disabled child.138 The result of this may be direct discrimination based on three different grounds: ethnicity, gender and disability. In those situations, the question arises as to which of the various specific exemption clauses applies: those related to disability, to gender or to ethnicity based discrimination.139 Although it is sometimes argued that in these cases the clause must be applied that offers "the highest degree of protection", exemption clauses usually lay down a particular balance of interests (*e.g.* between the prohibition of discrimination and the rights and interests of an employer or service provider). It is then sometimes difficult to know whose interests must be protected to the highest degree—those of the victim of unequal treatment, or those of the company or person responsible for it. Accordingly, it may be difficult to make a good choice for applying one or the other exemption clause.140

The problems in relation to intersectional discrimination are even more complicated.141 For example, if a person is disadvantaged because of her being a lesbian Muslim, she may suffer a type of highly specific and particularized discrimination that is more complex than a simple addition of the grounds of sexual orientation and religion.142

Closed systems can hardly deal with the specificities of such intersectional forms of discrimination.143 After all, closed systems accept that a discrimination must be *either* based on sexual orientation *or* on religion, and only one of the sets of exemptions can be applied.144 If it is impossible to choose between the two grounds, because a discrimination is caused by an intricate combination of them, it is difficult to determine which exemption should be applied. Moreover, the legislature usually defines exemptions to allow for particular types of permitted unequal treatment on specific grounds.145 It may be true, for example, that the legislature allows for an exemption for organizations based on religious ethos who want to hire only people that adhere to the same faith.146 The legislature may never have considered, however, if this exemption is still fair, an employer does not so much discriminate based on religion only, but on his opinion that it is not desirable to hire a lesbian Muslim woman.147 In the end, closed systems of exemptions may therefore fail to offer necessary clarity and legal protection and may ignore the complex reality of social stereotyping.148

---

138. *See, e.g.*, Janneke H. Gerards, *Discrimination Grounds*, *in* CASES, MATERIALS AND TEXT ON NATIONAL, SUPRANATIONAL AND INTERNATIONAL NON-DISCRIMINATION LAW 33, 171 (Dagmar Schiek et al. eds., Hart 2007) [hereinafter *Discrimination Grounds*]; *See also* EUR. FUNDAMENTAL RTS. AGENCY & EUR. CT. HUM. RTS, *supra* note 47, at 60.

139. *E.g.*, Gerards, supra note 82, at 159. For the difficulties related to a closed system of exemptions, *see also* EUR. FUNDAMENTAL RTS. AGENCY & EUR. CT. HUM. RTS, *supra* note 48, at 62.

140. Under the EU Directives, an additional difficulty is that the material scope of application of each Directive is different. *See* EUR. FUNDAMENTAL RTS. AGENCY & EUR. CT. HUM. RTS, *supra* note 48, at 62.

141. *See, e.g.*, Altman, *supra* note 107, §7.

142. *E.g.*, Shoben, *supra* note 137; Altman, *supra* note 107; EUR. FUNDAMENTAL RTS. AGENCY & EUR. CT. HUM. RTS., *supra* note 47, at 60.

143. This is even more true in the EU Non-Discrimination Directives, many of which are limited to a number of particular grounds, such as *either* gender *or* race/ethnicity; *see, e.g.*, *Discrimination Grounds*, *supra* note 138, at 172.

144. *Id.*

### 3. Avoidance Strategies

In the above, we have shown that there can be mismatches between a closed or semi-closed system and the need for protection against (multiple or intersectional) discrimination in a rapidly evolving society. These mismatches may result in a tendency with private actors or government bodies to rely on avoidance strategies. Companies, for example, can easily resort to such avoidance strategies because almost all closed systems only relate to direct discrimination, that is, discrimination that is demonstrably and decisively based on the listed grounds, such as ethnicity, disability, or gender. If the exemption clauses do not or no longer fit societal or employer needs, a company can be inclined to avoid direct forms of discrimination and hide them behind more neutral grounds or proxies.149

For example, if an employer does not want to hire any persons wearing religious clothing, he may obfuscate this direct religion-based discrimination by relying on neutrally formulated dress-codes that prohibit wearing head-covering clothing in the interest of "safety and hygiene" or the need to project a neutral company image.150 Even if it is clear that such dress codes cause *indirect*

---

145. This is clearly true for the EU Non-Discrimination Directives with their closed system of exemptions, but problematic examples also can be found in national law; *see id.*

146. On this debate, *see, e.g.*, Titia Loenen, *The Headscarf Debate. Approaching the Intersection of Sex, Religion and Race under the European Convention on Human Rights and EC Equality Law*, EUROPEAN UNION NON-DISCRIMINATION LAW. COMPARATIVE PERSPECTIVES ON MULTIDIMENSIONAL EQUALITY LAW 313 (Dagmar Schiek & Victoria Chege eds., Routledge, 2008).

147. *See, e.g.*, Fredman, *supra* note 137, at 16.

148. *See, e.g.*, EUR. FUNDAMENTAL RTS. AGENCY & EUR. CT. HUM. RTS., *supra* note 47, at 62; Makkonen, *supra* note 137, at 36; *Discrimination Grounds*, *supra* note 138.

149. Gerards, *supra* note 81, at 141.

150. *See* Case C-157/15, Achbita v. G4S Secure Solutions NV, ECLI:EU:C:2017:203 (Mar. 14, 2017). *See, e.g.*, Gerards, *supra* note 57, § 2.3 (Showing much debate on whether this constitutes direct or indirect discrimination). For further criticism of the indirect discrimination based on religion, the effect is that the detailed, closed list of exemptions do not apply.[151] Instead, the employer can make use of the open-formulated general exception for indirect discrimination.

In similar vein, courts sometimes willingly use the strategy of constructing indirect discrimination to avoid the lack of any specific exemptions that they consider necessary in the case they have at hand, yet have not been foreseen by the legislature when drafting the closed list of exemptions.[152] A well-known example in the Netherlands is that sexually active homosexuals were excluded from donating blood because of the heightened risk for HIV infection.[153] The reason for this exclusion was to protect public health, which at the time arguably could only be safeguarded by means of this exclusion.

However, the Dutch equal treatment legislation did (and does) not provide for any exemption on public health grounds. The then Equal Treatment Commission found a way out of this conundrum by construing a case of indirect discrimination. Although the rule that homosexuals were not allowed to donate blood could be qualified as constituting direct discrimination based on sexual orientation, the Equal Treatment Commission decided that, in fact, the difference in treatment was based on sexual behavior and therefore constituted *indirect* discrimination on this ground.[154] This allowed the commission to accept the justification advanced by the

---

discrimination approach taken in the *Achbita v. G4S Secure Solutions NV* case, *see, e.g.*, Eva Brems, *European Court of Justice Allows Bans on Religious Dress in the Workplace*, IACL-AIDC BLOG (Mar. 26, 2017), https://blog-iacl-aidc.org/test-3/2018/5/26/analysis-european-court-of-justice-allows-bans-on-religious-dress-in-the-workplace [https://perma.cc/ZA5Z-6WNC]; Titia Loenen, *In Search of an EU Approach to Headscarf Bans: Where to Go After Achbita and Bougnaoui?*, 10 REV. EUR. ADMIN. L., no. 2, 2017, at 47; Eleanor Spaventa, *What is the Point of Minimum Harmonization of Fundamental Rights? Some Further Reflections on the Achbita Case*, EU L. ANALYSIS BLOG (Mar. 21, 2017), http://eulawanalysis.blogspot.com/2017/03/what-is-point-of-minimum-harmonization.html [https://perma.cc/3PEX-8QQU]. The approach was received more favorably by Mark Bell, *Leaving Religion at the Door? The European Court of Justice and Religious Symbols in the Workplace*, 17 HUM. RTS. L. REV. 784, 792 (2017).

151. *See also* Heringa, *supra* note 81 at 28.
152. Gerards, *supra* note 81, at 138.
153. *See* Rotterdam, et al. v. Red Cross Blood Bank Central Netherlands et. al., 137, (Dutch Equal Treatment Commission 137 1998), https://www.mensenrechten.nl [https://perma.cc/V2KS-KRZ3]; *see Excluding Blood Donorship of Men*, 85, (Dutch Equal Treatment Commission 2007), https://www.mensenrechten.nl [https://perma.cc/V2KS-KRZ3]; Gerards, *Nieuwe ronde, nieuwe kansen: naar een 'semi-open' systeem van gelijkebehandelingswetgeving?*, *supra* note 81, at 144. A similar issue has come before the CJEU (but in that case, none of the EU Directives was applied, but Charter art. 21 was relied on); *see* Case C-528/13, Geoffrey Léger v. Ministre des Affaires sociales, de la Santé et des Droits des femmes, ECLI:EU:C:2015:288 (Apr. 29, 2015).
154. Rotterdam, *supra* note 153.

government under the open justification clause, and thereby escape the harshness and rigidity of the close system of exemptions for direct discrimination.155

Avoidance strategies such as these have the advantage of creating some degree of flexibility, but the need to rely on them also highlights some weaknesses of closed systems.156 Closed systems will not always help to fight direct discrimination on protected grounds, since direct discrimination can be hidden behind neutral grounds that allow for more open forms of justification. In addition, legal certainty and clarity are reduced if companies or courts resort to indirect discrimination to avoid the application of the limited system of exemptions.

## C. *Choice of Grounds in Closed and Semi-closed Systems*

In closed and semi-closed systems, it matters which grounds are protected. In fully closed models, the grounds matter because the exemptions are narrowly tailored to fit the specific characteristics of these particular grounds.

In cases on indirect discrimination too, it matters which grounds are protected by law. If somebody brings a discrimination case to court and provides "facts from which it may be presumed that there has been direct or indirect discrimination, it shall be for the respondent to prove that there has been no breach of the principle of equal treatment."157 EU non-discrimination law thus reverses the burden of proof once a case of indirect discrimination has been shown. Nevertheless, the complainant must still show a *prima facie* case of indirect discrimination, which requires that some relevant link with the protected ground is demonstrated. Moreover, we have shown that in systems with semi-closed lists or open possibilities of justification, the lists of grounds matter because of their symbolic and benchmark value.

As explained in Section IV(B)(1), we may come to realize over time that new grounds should be considered worthy of special protection as a consequence of societal or technological developments. In closed and semi-closed systems this may raise the question of whether and when any new grounds should be added to already existing lists of grounds. This question is more important if a closed list is considered fully exclusionary, meaning that no protection is given to discrimination based on non-listed grounds. In that case, the difference between protected and non-protected grounds is significant. Even if a list is semi-closed, however, the benchmark function bears a certain potential for *a contrario* reasoning: The fact that some grounds are expressly included, and others are not, can evoke the image that the non-listed grounds are less suspect and therefore less worthy of (special) protection or intensive judicial scrutiny.

---

155. *Id.*
156. *See* Gerards, *Nieuwe ronde, nieuwe kansen: naar een 'semi-open' systeem van gelijkebehandelingswetgeving? [New round, new opportunities: towards a 'semi-open' system of equal treatment legislation?]*, *supra* note 81, at 144.
157. Council Directive 2000/43, *supra* note 31, at art. 8 ¶ 1.

For those reasons, it should be carefully considered which grounds are included in closed or semi-closed systems and why. As we briefly discussed in Section III(E), there may be various reasons to include certain grounds in a non-discrimination provision. We mentioned, for example, the immutability of certain personal characteristics, the existence of deeply rooted stigmata and the risk for overbroad stereotyping. Because of the importance of the inclusion of certain grounds in non-discrimination law, we now deal with these reasons in more detail.

1. Choice and Non-choice Grounds

First, we discuss the rationale of immutability or "non-choice grounds" Unequal treatment based on characteristics people cannot help having (such as ethnicity or gender) is generally considered unfair.[158] By contrast, if people have a certain responsibility or choice to influence a ground for differential treatment, it is more readily accepted that it can reasonably be considered in decision-making. For example, the European Court of Human Rights has considered immigration status to be a "choice ground", since most individuals can freely decide whether they want to migrate to a different country to find a better life there.[159] For that reason, the Court has held that unequal treatment in social security measures that are based on immigration status are less suspect (and therefore more easily justifiable) than, for example, unequal treatment based on nationality, since the Court considers nationality as a status is relatively difficult to change and therefore is not a "choice ground".[160]

---

158. *See, e.g.*, Altman, *supra* note 107; *see also, e.g.*, Dagmar Schiek, *A New Framework on Equal Treatment of Persons in EC Law? Directives 2000/43/EC, 2000/78/EC and 2002/58/EC Changing Directive 76/207/EEC in Context*, 8 EUR. L.J. 290, 309 (2002).

159. Bah v. United Kingdom, App. No. 56328/07, ¶ 45 (Sept. 27, 2011), http://hudoc.echr.coe.int/eng?i=001-106448 [https://perma.cc/PZM4-YLN3]; *see also* Gerards, *supra* note 96, at 109; on the notion of choice grounds, *see also* MARIANNE H.S. GIJZEN, SELECTED ISSUES IN EQUAL TREATMENT LAW: A MULTI-LAYERED COMPARISON OF EUROPEAN, ENGLISH AND DUTCH LAW 285 (2006); *see also* Schiek, *supra* note 158, at 310.

160. *Bah*, App. No. 56328/07 at ¶ 45.

Although the distinction between "choice grounds" and "non-choice grounds" has a certain intuitive appeal, its use in drafting lists of grounds can be problematic.[161] Many non-choice grounds are regarded as a perfectly reasonable basis for (certain types of) decision-making in our societies, such as intelligence and talent (which may be used in admission to universities or music schools), or having a serious mental disorder (which may be used to exclude some persons from the right to vote or adopt a child).[162] Conversely, in western democracies, some grounds are considered to be hardly acceptable as a basis for differentiation, such as religion or belief. However, at least theoretically, an individual could consciously decide to abandon or change her faith. Apparently, thus, there is more to these grounds of discrimination than the pure immutability or innate nature of the characteristics they relate to.

The explanation for the problematic nature of the notion of choice and non-choice grounds can be found in the relation between this notion and the notion of core fundamental rights.[163] For example, within the freedom of religion, a distinction can be made between the *forum internum* (e.g. *holding* a certain religion or belief) and the *forum externum* (e.g. *expressing* a certain religion or belief).[164] The *forum internum*—the freedom of conscience, the freedom to hold certain opinions, one's personal (or gender, sexual or ethnic) identity—is usually protected in human rights documents as an absolute right that should not be restricted in any way, or as the core of a fundamental right.[165] One's conscience should be completely free since it is closely related to human dignity, liberty and personal autonomy, and it cannot reasonably be expected that a person change her conscience.[166] Clearly, to the extent that discrimination is based on a core element of one's person or identity, it cannot be reasonably regarded as based on a "choice ground", and discrimination based on these elements can be prohibited.

---

161. *See, e.g.*, *Altman, supra* note 107.

162. *See, e.g.*, JOHN H. ELY, DEMOCRACY AND DISTRUST. A THEORY OF JUDICIAL REVIEW 150 (1980).

163. *See* SANDRA FREDMAN, DISCRIMINATION LAW (2011); *see also Human Rights Transformed Positive Rights and Positive Duties* https://www.corteidh.or.cr/tablas/25653.pdf [https://perma.cc/26MM-77GR] (last visited Sept. 12, 2021).

164. *See, e.g.*, Pilvi Rämä, Towards the Core of Conscience. A Study on Conscientious Objection in the Workplace Under the Freedom of Thought, Conscience and Religion (2019) (Ph.D. dissertation, European University Institute) (unpublished, on file with the author). This distinction has been particularly well-established in the caselaw of the European Court of Human Rights; *see, e.g.*, CAROLYN EVANS, FREEDOM OF RELIGION UNDER THE EUROPEAN CONVENTION ON HUMAN RIGHTS 76 (2001); Peter Petkoff, *Forum Internum and Forum Externum in Canon Law and Public International Law with a Particular Reference to the Jurisprudence of the European Court of Human Rights*, 7 RELIGION & HUM. RTS. 183, 184 (2012). For the application of these notions to the context of choice/non-choice grounds, *see* Gijzen *supra* note 159, at 285; *see also* ROBERT WINTEMUTE, SEXUAL ORIENTATION AND HUMAN RIGHTS 177 (Oxford Univ. Press 1997).

165. Rämä, *supra* note 164; *see also Freedom of Religion*, CSW, https://www.csw.org.uk/freedomofreligion.htm (last visited Sept. 12, 2021).

By contrast, other forms of discrimination may affect the *forum externum*, which relates to the way in which someone gives expression to a certain religion, belief, opinion, or identity.[167] If the expressions are close to the core of an individual right (the *forum internum*), it may still be difficult to accept that a differentiation can reasonably be based on it; one can think here of same-sex partners living together, or believers wanting to attend a religious ceremony.[168] The further removed the expressions are from the core of fundamental rights, however, the more leeway can be seen to exist for balancing the right to non-discrimination to other fundamental rights or societal interests.

Indeed, in caselaw of the ECtHR and the CJEU, it can be seen that the right to wear religious clothing (such as headscarves) is balanced against the interest of interpersonal communication or neutrality in the workplace, or the right to voice one's political opinions is balanced against the interest of religious persons not to be offended because of their religious tenets.[169] Those aspects are less closely related to the absolute core of the freedom of religion or opinion and they therefore are also less absolutely protected as grounds of discrimination.

At first glance, the distinction between the *forum internum* and the *forum externum* seems to be a sharp one, just like the distinction between choice grounds and non-choice grounds. The above also shows, however, that in fact there is a spectrum that ranges from the core of grounds such as religion or belief to its periphery or fringes, and in which it can be more or less easily expected of people to freely make certain choices.170 This makes it difficult to use the criteria of "choice grounds" or "immutability" in closed or semi-closed lists of grounds. After all, it will often remain unclear exactly which aspects of these grounds can reasonably considered to be immutable and must be fully excluded as a fair basis for decision-making.171 Surely some such nuances can be taken into account in drafting a system of exemptions, but in many cases, in drafting non-discrimination legislation, an open possibility of justification will have to be relied on to do justice to the variety of different aspects that can be related to identity, religion, or political conviction. Indeed, this explains why the avoidance strategies discussed in Section IV(B)(3) are so often used in closed systems and in relation to the grounds mentioned.

---

166. *See, e.g.*, Eweida v. United Kingdom, App. No. 48420/10, 36516/10, 51671/10, 59842/10, ¶ 80 (Jan. 15, 2013), http://hudoc.echr.coe.int/eng?i=001-115881[https://perma.cc/8A9K-7WKJ]; *see also* Kokkinakis v. Greece, App. No. 14307/88, ¶ 33 (May 25, 1993), http://hudoc.echr.coe.int/eng?i=001-57827 [https://perma.cc/B3BY-2TMD]; *see e.g.*, MARIANNE H.S. GIJZEN, SELECTED ISSUES IN EQUAL TREATMENT LAW: A MULTI-LAYERED COMPARISON OF EUROPEAN, ENGLISH AND DUTCH LAW 285 (2006); *see also, e.g.*, Schiek, *supra* note 158, at 310.

167. *E.g.*, Rämä, *supra* note 164; Evans, *supra* note 164; Petkoff, *supra* note 164; Gijzen, *supra* note 159; Wintemute, *supra* note 164.

168. *See* Rämä, *supra* note 164; Gerards, Nieuwe ronde, nieuwe kansen: naar een 'semi-open' systeem van gelijkebehandelingswetgeving? *[New Round, New Chances: Toward a 'Semi-Open' System of Equal Treatment Law?]*, *supra* note 82, at 147; for a further discussion of different types of religious expression, *see e.g.*, Javier Martínez-Torrón, *Manifestations of Religion or Belief in the Case Law of the European Court of Human Rights*, 12 RELIGION & HUM. RTS. 112, 120–27 (2017).

169. On headscarves, *see, e.g.*, Eweida, App. No. 48420/10, 36516/10, 51671/10, 59842/10, ¶ 80; Achbita, ECLI:EU:C:2017:203; Case C-188/15, Bougnaoui v. Micropole SA, ECLI:EU:C:2017:204 (Mar. 14, 2017); *see, e.g.*, E.S. v. Austria, App. No. 38450/12 (Oct. 25, 2018), http://hudoc.echr.coe.int/eng?i=001-187188 [https://perma.cc/JNR8-BWYA].

### 2. Prejudice, Stereotyping, Stigmatization, and Vulnerability

Because of the problems related to the immutability rationale, legal scholars often hold that factors such as the immutability or responsibility for certain personal characteristics or expressions thereof do not constitute an appropriate basis for including these characteristics in (semi-)closed lists.172 Instead, scholars usually prefer another rationale, which is the perceived irrelevance of certain personal characteristics or behavior as a basis for decision-making or acting, or the need to protect certain groups against particular disadvantages.173 Implicitly, most non-discrimination systems are based on the presumption that important societal goods (such as access to employment or to certain services) should be distributed on rational grounds.174 Irrational considerations, such as prejudice, overbroad stereotypes or societal stigmata, are not allowed to have any tangible influence on the decisions made by governmental bodies, companies or employers.[175]

---

170. Gerards, *supra* note 81, at 147.
171. *See supra* text accompanying note 81.
172. *See, e.g.*, Altman, *supra* note 107.
173. *See, e.g.*, Alexandra Timmer, *Toward an Anti-Stereotyping Approach for the European Court of Human Rights*, 11 HUM. RTS. L. REV. 707 (2011).
174. For this and other rationales, *see* Altman, *supra* note 107; *see also, e.g.*, Christopher McCrudden & Haris Kountouros, *Human Rights and European Equality Law*, *in* EQUALITY LAW IN AN ENLARGED EUROPEAN UNION. UNDERSTANDING THE ARTICLE 13 DIRECTIVES 73, 74 (Helen Meenan ed., Cambridge University Press 2007).

In addition, non-discrimination legislation can be informed by the desire to provide vulnerable groups with additional protection against discrimination.[176] The members of vulnerable groups share characteristics that make that they, as a group, have long been subjected to discrimination and continue to suffer the results thereof, or are likely to suffer societal stigmatization or exclusion from important social goods or political decision-making.[177]

From a theoretical perspective, reliance on the rationales of irrelevance, stereotyping, stigmatization, prejudice, and vulnerability may be preferred over the immutability rationale for including grounds in systems of non-discrimination.[178] Nevertheless, the stereotyping rationale has some weaknesses of its own. As we noted in Section IV(A)(1), societal opinions are subject to change, which can translate in the gradual recognition of (new) patterns of stigmatization or exclusion. The slowness of this recognition may go paired with controversy over whether structural or systematic discrimination or overbroad stereotypes really exist in relation to certain groups. Currently, for example, it is debated whether the stereotyping rationale could apply to persons with lower education, low-income groups, persons characterized by an unhealthy lifestyle, persons with unfavorable genotypes, obese persons, or asylum seekers.[179] In this twilight zone it may be difficult for legislators to decide whether they should add these characteristics to closed or semi-closed lists. This also puts the symbolic value of closed or semi-closed lists into perspective: Although the listed grounds have a certain benchmark value and allow for new grounds to be compared to grounds that are already prohibited, there will always be debate about grounds that *just* fall outside the list and therefore seem to be granted less protection.

---

175. *See, e.g.*, Altman, *supra* note 107; *see also* TIMO MAKKONEN, MULTIPLE, COMPOUND AND INTERSECTIONAL DISCRIMINATION: BRINGING THE EXPERIENCES OF THE MOST MARGINALIZED TO THE FORE 2 (2002).

176. *See, e.g.*, Alexandra Timmer, *A Quiet Revolution: Vulnerability in the European Court of Human Rights*, *in* VULNERABILITY: REFLECTIONS ON A NEW ETHICAL FOUNDATION FOR LAW AND POLITICS 147 (Martha A. Fineman & Anna Grear eds., Ashgate 2013); Lourdes Peroni & Alexandra Timmer, *Vulnerable Groups: The Promise of an Emerging Concept in European Human Rights Convention Law*, 11 INT'L J. CONST. L. 1056 (2013); PROTECTING VULNERABLE GROUPS: THE EUROPEAN HUMAN RIGHTS FRAMEWORK (Francesca Ippolito & Sara Iglesias Sánchez eds., Hart 2015); INGRID NIFOSI-SUTTON, THE PROTECTION OF VULNERABLE GROUPS UNDER INTERNATIONAL HUMAN RIGHTS LAW (Routledge 2017); Valeska David, Crossing Divides and Seeking the Whole: An Integrated View of Cultural Difference and Economic Disadvantage in Regional Human Rights Courts (2018) (Ph.D. dissertation, Ghent University) (on file with the Ghent University Library).

177. *Cf.* Martha A. Fineman, *Equality, Autonomy, and the Vulnerable Subject in Law and Politics*, VULNERABILITY: REFLECTIONS ON A NEW ETHICAL FOUNDATION FOR LAW AND POLITICS 20-21 (Martha A. Fineman & Anna Grear eds., Ashgate 2013) (showing a different, more individualised type of conceptualization); *see also* Ippolito, *supra* note 38.

178. *See, e.g.*, Altman, *supra* note 107.

179. *See generally* Gerards, *supra* note 51; *see also, e.g.*, Lavrysen, *supra* note 39 (showing debates on whether low-income groups are stereotyped).

### 3. The Flexibility of Open Lists

For fully open lists, all of the above is different, since they do not contain any grounds at all. Instead they give full flexibility to courts to decide in individual cases whether there is a suspect case of discrimination and whether compelling reasons should be asked as justification.[180] In making their decisions, courts can take account of all factors and rationales discussed above, ranging from the individual's own responsibility and the closeness of a certain characteristic, to the core of a fundamental right such as the freedom of expression and societal changes in views on the rationality of a certain characteristic as a basis for decision-making.[181]

However, recurring problems in open systems are the risk of judicial bias and the lack of legal certainty.[182] As is often the case with lawmaking, nuance comes at the cost of clarity. In the end, therefore, choosing a closed, semi-closed, or fully open system requires careful analysis of advantages and disadvantages.

### D. *The Difference Between Unequal Treatment and Non-discrimination*

When deciding for a certain type of non-discrimination law, one last factor that must be considered concerns the difference between equal treatment and non-discrimination.[183] The principle of non-discrimination mainly aims to ban unequal treatment based on suspect grounds and fight the underlying irrational or irrelevant considerations for decision-making.[184] In a non-discrimination rationale, unequal treatment based on suspect characteristics is *a priori* wrong and unacceptable.[185] This presumption of unacceptability can be rebutted only if it can be demonstrated that the difference in treatment is not really based on the suspect characteristics, but on objective reasons that are fully rational, neutral, and fair.[186] One may think of examples of limiting breast cancer prevention treatment to women or setting language requirements for a language teacher.[187]

---

180. Gerards, *supra* note 81, at 153.
181. *See supra* text accompanying note 81.
182. *See also* SANDRA FREDMAN, DISCRIMINATION LAW (2d ed., Oxford Univ. Press 2011); Christoph K. Winter, *The Value of Behavioral Economics for EU Judicial Decision-Making*, 21 GER. L.J. 240, 241 (2019).
183. This section is based on Gerards, *supra* note 96.
184. *Cf., e.g.*, Rory O'Connell, *Cinderella Comes to the Ball: Art 14 and the Right to Non-Discrimination in the ECHR*, 29 LEGAL STUD. 211 (2009).
185. *See Gerards supra* note 96, at 114.

The non-discrimination rationale also leaves room to contest indirect forms of discrimination as long as a concrete case can be proven of clear and structural societal disadvantaging of already stigmatized, vulnerable, or stereotyped individuals.[188] The non-discrimination rationale thus reflects a clear human rights approach towards unequal treatment, as the rationale is based on notions of human dignity, personal autonomy, fair and rational distribution of important social goods, and protection of vulnerable and neglected social groups.[189]

By contrast, the general principle of equality is a rather empty legal principle with no moral content of its own.[190] "The function of the equality principle is mainly instrumental or procedural."[191] The equality principle can help an applicant demonstrate arbitrariness or unfair treatment by pointing at other cases, similar to her own, where more favorable decisions have been taken or which are governed by a less burdensome legal regime. Hence, if an equal treatment perspective is taken, each difference in treatment that affects an applicant's rights should be assessed by a court for reasonableness and fairness. Following the equal treatment rationale, the ground on which the difference in treatment is based is not relevant to the applicability of a test of justification.[192] The only relevant question is if one group or person is allowed to exercise a certain right or receive a certain benefit, whilst this is not permitted for another person or group.[193] The equal treatment rationale thus differs from the non-discrimination approach, which has more normative content of its own.[194]

---

186. *See* Altman, *supra* note 107; *see also* McCrudden, *supra* note 174, at 75.
187. Gerards, *supra* note 96, at 115.
188. This notion of structural disadvantage is usually behind the acceptance of the concept of indirect discrimination. *See, e.g.*, Altman, *supra* note 107, § 4.2 (arguing that indirect discrimination can also be contested if underprivileged individuals are targeted).
189. Samantha Besson, *Gender Discrimination Under EU and ECHR Law: Never Shall the Twain Meet?*, 8 HUM. RTS. L. REV. 647, 653–54 (2008).
190. *See* ALF ROSS, ON LAW AND JUSTICE Jakob v. H. Holtermann ed., Uta Bindreiter transl., (2019); *See* John R. Lucas, *Against Equality*, 40 PHIL. 296 (1965); *See* Peter Westen, *The Empty Idea of Equality*, 95 HARV. L. REV. 537, 547 (1982); *See* Erwin Chemerinsky, *In Defense of Equality: A Reply to Professor Westen*, 81 MICH. L. REV. 575, 582 (1983).
191. Gerards, *supra* note 96, at 117.
192. The ground of discrimination may certainly be relevant to the intensity of review and the applicability of the very weighty reasons test, but this is a different matter than the justiciability of a difference in treatment. EDEL, *see supra* note, 95, 117–18 and accompanying text.

Moreover, different from the non-discrimination rationale, the equal treatment principle implies that all forms of different treatment can be subjected to judicial review, regardless of the relevant ground, including seemingly futile cases, such as cases about a slight difference in taxation for owners of different types of cars.[195] For all such cases, courts need to examine whether the alleged differences in treatment constitute a disadvantage that can be reasonably justified. This could be considered burdensome and require more judicial capacity than is actually called for by the relative unimportance of such cases.[196]

### E. Which Way to go?

As we discussed in Section III, there are three main systems of non-discrimination law: closed, open, and hybrid systems (which are in-between options as regards their lists of grounds and/or systems of exemptions). In Section IV, we explained that all these conceivable systems have their own strengths and weaknesses.[197] Systems based on semi-closed lists of grounds provide for relatively clear and predictable benchmarks to show which grounds of discrimination are suspect, and semi-closed lists of exemptions provide for considerable legal certainty and clarity. Fully closed systems of grounds and exemptions allow for stronger democratic legitimacy than open systems do, since the latter mainly leave it to courts to decide which forms of unequal treatment are problematic and need to be prohibited. Closed systems also protect courts from being burdened by unimportant, futile cases about all thinkable forms of unequal treatment.

By contrast, fully open systems have the advantage of being flexible. They leave leeway for acknowledging hidden forms of discrimination, for protecting multiple and intersectional forms of discrimination, and for accepting new grounds of discrimination as suspect. The fact that these open systems also allow seemingly futile cases to be brought thereby has a value of its own, which is that such cases may allow courts to incrementally discover patterns of discrimination that are more problematic and suspect than is often thought.

---

193. This comes down to a test of (comparative) disadvantage; *See* JANNEKE H. GERARDS, JUDICIAL REVIEW IN EQUAL TREATMENT CASES, 77–78; and, for the desirability of its application instead of the classic test of comparability and its practicability, 669–75 (Martinus Nijhoff/Kluwer Law International, 2005).

194. *See* Kent Greenawalt, *How Empty is the Idea of Equality?*, 83 COLUM. L. REV. 1167 (1983); Erwin Chemerinsky, *In Defense of Equality: A Reply to Professor Westen*, 81 MICH. L. REV. 575 (1983).

195. Gerards, *supra* note 81, at 154.

196. Gerards, *see supra* note 81, 155 and accompanying text..

197. *See supra* note 85, 166 and accompanying text; *see also supra* note 163.

This complex set of characteristics, strengths, and weaknesses makes it difficult to choose one or the other type of system. However, the current section has also demonstrated that it may not be necessary to choose, since a hybrid system might offer a good way out.[198] In particular, in theory, there is considerable advantage in combining a semi-closed list of grounds with a semi-closed system of exemptions.[199] The listed grounds and exemptions in such a system may act as benchmarks to give some indication as to the type of characteristics that are generally considered to be suspect and problematic as a basis for decision-making and to the kind of exemptions that generally can be made. These indications help to create a certain degree of legal certainty and make it easier for courts to know how they can build on the lists provided.

In addition, in such a combined system, the democratically legitimized legislature will play an important role in defining the grounds to be put on the list and the semi-closed exemptions. This democratic influence will add to the perceived legitimacy of non-discrimination law and will allow for societal and political debates to be reflected in the relevant constitutional or legislative provisions.

Simultaneously, the relative openness of such a combined system allows for flexibility and adaptation over time. If new or more nuanced exemptions are needed, courts do not necessarily have to rely on notions of indirect discrimination (with all inherent problems related to the burden of proof), but they may directly require a reasonable justification to be advanced. This flexibility in applying exemptions allows courts to keep an open eye for intersectional, multiple, and proxy-based discrimination and ask for a reasonable justification for such differences in treatment.[200] In addition, courts could use the leeway left by semi-closed grounds to add new (aspects of) grounds to meet societal and technological developments and to adequately respond to a growing recognition of the suspectness of certain grounds of discrimination.

Hence, in theory, we submit that it would be most valuable to opt for a hybrid system of non-discrimination law, whereby a semi-closed list of grounds and a semi-closed system of exemptions are

---

198. Gerards, *supra* note 81, at 144, 166.
199. Gerards, *supra* note 81, at 144.
200. Gerards, *supra* note 81, at 166.

combined.201 In the EU non-discrimination laws discussed in Section II, this particular hybrid does not exist, which means that introducing it will require a strong legislative effort.

## V. Systems of Non-discrimination Law in Relation to AI

### A. *In Brief: Specifics of AI-driven Discrimination and Systems of Non-discrimination Law*

We have found in Section IV that in theory, a hybrid system of non-discrimination law, with a semi-closed list of grounds and a semi-closed set of exemptions, makes the most of the advantages of the different theoretical systems and reduces their weaknesses to the greatest possible extent. This leaves us with the question of whether the same conclusion can be reached in relation to what system of non-discrimination law can best be applied to AI-driven discrimination, considering the special challenges that AI presents us with? We explained these special challenges in Section I. To summarize, we distinguished two categories of problems with AI-driven discrimination: (i) problems related to harm done to people with protected characteristics, such as ethnicity, and (ii) problems related to differentiation that does not necessarily harm people with protected characteristics, but that is still unfair.

In relation to category (i), we set out that AI is good at identifying correlations between factors that perhaps are not identical to ethnicity, but (especially if seen in combination) may come very close: proxy discrimination. Proxy discrimination is often unintentional and can be an effect of societal and structural factors, such as concentrations of persons with specific characteristics (such as a certain ethnicity) living in the same neighborhoods. In such situations, some kind of legal creativity is necessary to show that an AI-driven case of differentiation based on proxies amounts to a prohibited case of discrimination on a particular ground of discrimination.

In relation to category (ii), we showed that AI-driven differentiation may lead to discrimination on grounds that are not traditionally regarded as suspect grounds, such as postal codes or the type of car people drive. Nevertheless, if premiums for car insurance are based on postal codes, or credit is refused because of the type of car a person drives, these grounds may be perceived by individuals as irrelevant and they may regard the resulting unequal

---

201. *Id.* at 129. Critically on the current set-up of the EU non-discrimination directives. *See generally* Dagmar Schiek, *Broadening the Scope and the Norms of EU Gender Equality Law: Towards a Multidimensional Conception of Equality Law*, 12 MAASTRICHT J. EUR. & COMP. L. 427 (2005).

NON-DISCRIMINATION IN AI                                                6/1/22  12:55 PM

50                            COLO. TECH. L.J.                        [Vol. 20.1treatment as unfair. In addition, unequal treatment based on such specific factors as postal code, willingness to pay certain prices for certain goods or services, or car brands may cause structural or systemic disadvantages for certain groups in society, such as low-income groups.

As we demonstrated in Section III, in considering which system of non-discrimination law would best be suited to dealing with AI-driven discrimination,[202] choices must be made as to whether: (i) the non-discrimination legislation should have a closed, semi-closed, or open list of grounds and (ii) the legislation should have a closed, semi-closed or open list of exemptions. Making these choices is difficult, but in Section IV we listed a number of characteristics of the various systems and subsystems that might help us decide. Taking the various strands together, we now explore the question what system would be the preferred option if we take account of the challenges set by AI-driven discrimination.

### B. *Determining the Preferred System of Non-discrimination law to Deal with AI-driven Differentiation*

#### 1. Closed, Semi-closed, or Open List of Grounds of Discrimination?

We first address whether a closed, semi-closed or open list of grounds of discrimination would best fit the specific characteristics of AI-driven discrimination. We submit that a fully closed system of grounds would not help to fight unfair forms of AI-driven differentiation. It would be easy for organizations using AI to (intentionally or unintentionally) circumvent the exhaustively enumerated lists of grounds by focusing on seemingly harmless proxies. Clearly, in such cases of discrimination by proxy, it will be difficult (for victims or regulators for example) to identify, let alone prove that such a discrimination is directly based on one of an exhaustively listed number of grounds. The concept of indirect discrimination can help victims, since it allows victims to show that differentiation harms a group with a protected characteristic. Even in cases of indirect discrimination, however, victims must prove that there is a *prima facie* case of indirect discrimination.[203] Proving such a *prima facie* case can be burdensome.

---

202. *See generally supra* notes 104–06 and accompanying text.
203. *See, e.g.*, EUROPEAN FUNDAMENTAL RIGHTS AGENCY & EUROPEAN COURT OF HUMAN RIGHTS, HANDBOOK ON EUROPEAN NON-DISCRIMINATION LAW 232 (2018 ed. 2018) (explaining the requirements needed to show *prima facie* of indirect discrimination).

A fully open system of grounds would fit the high degree of specificity of AI-driven differentiation much better than a fully closed list of grounds. However, as we explained in Section IV, a fully open list of grounds has disadvantages of its own. For instance, an open list creates little legal certainty for companies and other organizations regarding the proxies and factors that they can reasonably use as a basis for AI-driven decision-making. Moreover, in each case of AI-driven differentiation, companies could be required to show that there is a reasonable justification for the difference in treatment.

Consequently, we argue that AI-driven discrimination can best be tackled by means of non-discrimination legislation with a semi-closed list of grounds. First, as we explained in Section IV(A), such a semi-closed list can have an important benchmark function and can serve to emphasize the symbolic value of the nondiscrimination law. Companies, public bodies, and courts can compare the concrete ground or proxy before them to the listed grounds to see if they are somehow similar, for instance because they reflect biases, prejudice, or overbroad stereotyping in a comparable manner. If such similarities exist in a concrete case, the discriminator must advance weighty and convincing reasons in order to justify the use of certain proxies or factors as a fair and relevant ground for decision-making.

Second, an advantage of a semi-closed list is that it is not necessary for a court to connect (directly or indirectly) the concrete ground of AI-driven differentiation to one of the grounds specifically listed in the applicable nondiscrimination provision. This makes it possible to deal with the problem of differentiation on seemingly irrelevant criteria discussed in Section I(C)(2). For example, if an algorithm distinguishes on the basis of car brand or postal code, it is possible to challenge the resulting differentiation before a court directly, that is, without it being necessary to claim that, indirectly, the differentiation amounted to indirect discrimination on one of the listed grounds, such as gender or ethnic origin. This makes it easier for individuals to challenge AI-driven discrimination that seems irrelevant to them or that, in their view, affects them in an unfair manner.

Third, an advantage of a semi-closed list of grounds is that it combines important elements of an equal treatment rationale with the elements of a pure nondiscrimination rationale, as discussed in Section IV(D). A nondiscrimination system with a semi-closed list of grounds can provide an adequate response to the problem of incorrect predictions discussed in Section I(C)(2). In that section we showed that an algorithm might make an accurate prediction for an entire group, but this prediction may not be correct for a particular individual. Based on an equal treatment rationale, an individual could argue that this causes an unwarranted equal treatment of unequal cases for which a proper justification is required, regardless of the ground on which such substantive unequal treatment is based.

Fourth, a semi-closed list of grounds allows for the addition of new protected grounds by means of caselaw. This is important since we showed in Section I(C)(1) that AI-driven decision-making can lead to novel forms of structural disadvantage that are related to characteristics such as socioeconomic status and relative poverty. A semi-closed list of grounds allows for challenging concrete cases of disadvantage based on such characteristics in a court.[204] Over time, moreover, courts can come to recognize that groups defined by a low socioeconomic status or relative poverty suffer structural social and political disadvantages and can be typified as vulnerable groups. This might then lead courts to accept that (direct or indirect) discrimination based on income or socioeconomic status is suspect and therefore in need of very weighty reasons as justification. Eventually this may help tackle some of the structural social disadvantages that stem from AI-driven discrimination.

Hence, in our view, a semi-closed list of grounds could help address many of the particularities and effects of AI-driven differentiation. This means that nondiscrimination laws sporting a fully closed list of grounds, such as the EU Directives, are less ideal as legal bases to fight AI-driven discrimination. By contrast, Article 14 ECHR and Article 21 CFR both have a semi-closed list of grounds, and therefore, from the perspective of the "lists of grounds" constitute useful provisions to tackle AI-driven discrimination.

In sum, we found that AI-driven discrimination can best be tackled by non-discrimination law with a semi-closed list of grounds. This still leaves us with the question of whether a closed, semi-closed or open system of exemptions and justification should be preferred. We discuss this question next.

---

204. Provisions such as ECHR art. 14 and Charter art. 21 explicitly mention "property" and "social origin." The EU Fundamental Rights Agency and the European Court of Human Rights have derived from international law documents that the ground of "social origin" can relate to "one's social situation, such as poverty and homelessness." *See id.* at 216. Accordingly, there is a useful benchmark available in these semi-closed lists that could be used by courts to also deem income-based discrimination to be covered by these provisions.

### 1. Closed, Semi-closed or Open List of Exemptions?

As discussed in Section IV(E), from a theoretical perspective, a semi-closed system of exemptions is to be preferred over a fully open or fully closed system. Fully closed systems of exemptions create legal certainty; companies, public authorities, and courts only need to apply the exemptions as they are laid down in the legislation. In addition, we showed that clear legislative exemptions have the advantage of democratic legitimacy, since it is the democratically accountable legislature who is in charge of drafting and amending the exemptions.

However, we also explained that an important disadvantage of closed systems of exemptions is that they are rigid and that it will be difficult for a legislator to foresee all possible needs for exemptions. This is particularly problematic in relation to AI-driven discrimination. As discussed, AI allows for differentiation based on a wide number of different factors, characteristics and proxies, all of which may occur in an equally wide number of situations. AI-driven differentiations may be used in price differentiation by online stores, but also in predictive policing; AI may guide the hiring policy of a company, but it may also help diagnose specific illnesses or diseases. It would be nearly impossible to draft neat, precise, and future-proof exemptions for all such forms of differentiation and include all of them in one (or a few) clear pieces of general nondiscrimination legislation.

Indeed, the EU Directives illustrate the difficulties when designing future-proof nondiscrimination law. Each of the many different nondiscrimination Directives relate to specific fields of application (employment, provision of services, sometimes aspects of social security and social benefits), and are restricted to a number of exhaustively listed grounds. These Directives are more effective in fighting discrimination *because* of these restrictions of their material scope. If more grounds were covered, or more fields of application were included, this would necessitate adding more exemption clauses to accommodate the specific characteristics of—for example—predictive policing or online price differentiation. Alternatively, the exemptions could be drafted in wide-ranging terms, for instance allowing an alleged discriminator to invoke a reasonable justification. However, that would cause the exemptions to lose their distinctive value and would make them similar to the justifications allowed in open systems of exemptions.

Consequently, it is difficult to draft fully closed lists of exemptions for unequal treatment that are able to deal with the specific challenges of AI-driven differentiation. The same is true for semi-closed lists of exemptions. As we showed in Section IV, the advantage of semi-closed lists is that they can constitute a benchmark and make clear to users (companies, public authorities, courts, etc.) what kind of considerations can be regarded as acceptable justifications. In light of the wide range of situations in which AI-driven discrimination can occur and knowing the diverse types of differentiation that AI may cause, however, it will not be possible to come up with a comprehensive, understandable and useful set of benchmark exemptions that can easily be included in a general nondiscrimination law.

We therefore argue that, if the choice is made to deal with AI-driven differentiation by means of general nondiscrimination legislation, an open system of exemptions is the only solution that can be made to work. This means that for all forms of AI-driven differentiation between people, the organization using AI must be able to show a fair and reasonable justification. An AI-driven difference in treatment thus must be shown to be conducive and necessary to serving a legitimate aim; and there must be a reasonable relationship of proportionality between that aim and the disadvantage suffered by an individual or group. If challenged, it is up to organizations using AI to ensure such justifications can be advanced, and it must be possible for individuals to challenge them before supervisory bodies and courts. In turn, courts must be able to trace the relevant ground for differentiation in light of the benchmark grounds provided in the nondiscrimination act and must be able to set higher or lower standards for the justification to be met in line with the "suspectness" of the grounds used (directly or indirectly) in the case of AI-driven discrimination before it.

CONCLUSION

To conclude, AI that is used to make decisions about people challenges non-discrimination law in several ways. First, AI-driven decisions can harm people with protected characteristics such as ethnicity or gender. Second, AI can harm certain groups that are not protected under non-discrimination law. For example, AI could affect groups that are defined by many data points (car type, postal code, etc.) and have little to do with characteristics traditionally covered by non-discrimination law.

We think that a hybrid system of non-discrimination law, with a semi-closed list of grounds and an open possibility for exemptions and justification, is best-suited to deal with the particularities of AI-driven discrimination. A semi-closed list of grounds is an illustrative list of prohibited grounds of discrimination, typically ending with a phrase such as "or and other ground".

We are aware that such a hybrid system does not create a high level of legal certainty and it may be considered overly flexible. Nevertheless, such a system does allow for an adequate level of protection against individual and structural AI-driven discrimination, and it has a considerable symbolic and democratic value.

Insofar as European or national legislation is already in place that prohibits particular forms of discrimination and covers AI-driven discrimination in specific fields, it is important to retain such legislation. However, in light of our findings, we plead against introducing general equal treatment laws specifically pertaining to AI-driven discrimination, or merely adding new grounds or new exemptions to existing closed non-discrimination laws to accommodate for AI-driven discrimination. Such legal steps would not help to address the specific challenges following from AI-driven discrimination, and they would not fit well with the theoretical considerations for choosing a particular non-discrimination system.

Instead, (national) legislators should retain or introduce general non-discrimination provisions that can act as a safety net if specific, closed non-discrimination clauses do not apply or if they result in problematic avoidance strategies by companies and other organizations. To some extent, Article 14 ECHR and Article 21 EU Charter function as a safety net, but the limitations of those provisions should be noted: Article 14 ECHR only applies in relation to the exercise of other ECHR rights, whereas Article 21 EU Charter only applies in relation to the implementation of EU law.[205] In addition, due to constitutional constraints at the national level, individuals may not always be able to directly invoke Article 14 ECHR.

Therefore, national legislators should consider introducing a general discrimination clause of the same nature as these treaty provisions, in a national constitution or statute, which people can directly invoke before national courts, in relation to discrimination by both public authorities and private bodies. Introducing such a provision may be a major challenge, but from a perspective of being able to properly address AI-driven discrimination, it is certainly worth it.

---

205. *See* EUROPEAN FUNDAMENTAL RIGHTS AGENCY, *supra* note 204, at 23, 28.